%% file: sn-article.tex
\documentclass[pdflatex,sn-mathphys-ay]{sn-jnl}

\usepackage{graphicx}%
\RequirePackage{amsthm,amsmath,mathtools,multirow,amsfonts,bbm,amssymb,xcolor}
\usepackage{mathrsfs}%
\usepackage[title]{appendix}%
\usepackage{textcomp}%
\usepackage{manyfoot}%
\usepackage{booktabs}%
\usepackage{algorithm}%
\usepackage{algorithmicx}%
\usepackage{algpseudocode}%
\usepackage{listings}%

\usepackage{todonotes}
\usepackage{pdflscape}
\usepackage{pdfpages}
\usepackage{setspace} 
\usepackage{longtable}
\usepackage{float}
\hypersetup{
  colorlinks=true,
  citecolor=blue,
  linkcolor=blue,
  urlcolor=blue
}
\usepackage[defaultlines=2,all]{nowidow} 
\usepackage{enumitem} 
\usepackage{subfiles} 
\usepackage{changes} 
\usepackage[normalem]{ulem} 
\usepackage{cancel}
\usepackage{subfiles}

\definechangesauthor[name={Me}, color=red]{me}

\DeclareMathOperator{\KL}{KL}
\DeclareMathOperator{\JS}{JS}
\DeclareMathOperator*{\argmin}{arg\,min}
\DeclarePairedDelimiter\abs{\lvert}{\rvert}%
\newcommand{\JSGa}{\mathop{\mathrm{JS}^{\mathrm{G}_{\lambda}}}\nolimits}
\newcommand{\cJSGa}[1]{\mathop{\mathrm{cJS}_{i, #1}^{\mathrm{G}_{\lambda}}}\nolimits}
\newcommand{\eJSGa}[1]{\mathop{\mathrm{eJS}_{i, #1}^{\mathrm{G}_{\lambda}}}\nolimits}

\newcommand{\rhou}[1]{\rho_{\text{#1}}}
\newcommand\numberthis{\addtocounter{equation}{1}\tag{\theequation}}

\newcommand{\mvmean}[1]{%
\boldsymbol{\alpha}_{\mid i,#1}y
+
y^{\boldsymbol{\beta}_{\mid i,#1}}
\boldsymbol{\mu}_{\mid i,#1}
}

\theoremstyle{thmstyleone}%
\newtheorem{theorem}{Theorem}
\newtheorem{proposition}[theorem]{Proposition}%

\theoremstyle{thmstyletwo}%

\theoremstyle{thmstylethree}%
\usepackage{xr-hyper}
\begin{document}

\title[Article Title]{\bf Clustering of multivariate tail dependence using conditional methods}


\author*[1]{\fnm{Patrick} \sur{O'Toole}}\email{pot23@bath.ac.uk}

\author[1]{\fnm{Christian} \sur{Rohrbeck}}\email{cr777@bath.ac.uk}
\equalcont{These authors contributed equally to this work.}

\author[2]{\fnm{Jordan} \sur{Richards}}\email{Jordan.Richards@ed.ac.uk}
\equalcont{These authors contributed equally to this work.}

\affil*[1]{\orgdiv{Department of Mathematical Sciences}, \orgname{University of Bath}, \orgaddress{\city{Bath}, \country{UK}}}

\affil[2]{\orgdiv{School of Mathematics and Maxwell Institute for Mathematical Sciences}, \orgname{University of Edinburgh}, \orgaddress{\city{Edinburgh}, \country{United Kingdom}}}



\abstract{
  The conditional extremes (CE) framework has proven useful for analysing the joint tail behaviour of random vectors.
  However, when applied across many locations or variables, it can be difficult to interpret or compare the resulting extremal dependence structures, particularly for high-dimensional vectors. 
  To address this, we propose a novel clustering method for multivariate extremes using the CE framework.
  Our approach introduces a closed-form, computationally efficient dissimilarity measure for multivariate tails, based on the skew-geometric Jensen-Shannon divergence, and is applicable in arbitrary dimensions.
  Applying standard clustering algorithms to a matrix of pairwise distances, we obtain interpretable groups of random vectors with homogeneous tail dependence.
  Simulation studies demonstrate that our method outperforms existing approaches for clustering bivariate extremes under varied simulated dependence structures, and uniquely extends to the multivariate setting. 
  In our application to Irish meteorological data, our clustering identifies spatially coherent regions with similar extremal dependence between precipitation and wind speeds. 
}




\maketitle

\section{Introduction}\label{sec:intro}



Extreme Value Theory (EVT) provides asymptotically justified models for estimating the probability of events outside the range of observed values \citep{Coles2001}.
EVT has found broad usage in numerous, disparate domains, including finance \citep{Poon2004}, public health \citep{Vettori2019}, neuroscience \citep{Talento2025}, and ecology \citep{Koh2025}.
Environmental applications are one area which benefits from the use of EVT methods, to estimate the severity of future extreme events and develop risk management strategies. 
Examples include the analysis of heatwaves \citep{tanarhte2015heat, french2019quantifying}, extreme rainfall \citep{osei2021estimation}, wind speeds \citep{steinkohl2013extreme, soukissian2015effect}, wildfire risks \citep{Koh2023}, and the impact of environmental extremes on insurance claims \citep{Rohrbeck_flooding_2018, Miralles2024}.

In the multivariate EVT setting, one is interested in the \emph{joint} tail behaviour of a random vector.
Multivariate extremes frequently arise in spatial applications, where one or more variables are recorded at a discrete set of sampling locations.
Within such settings, simultaneous or ``concomitant'' extremes, where multiple variables or locations concurrently exhibit extremal behaviour, can produce substantially more severe outcomes than single-variable extremes \citep{zscheischler2018future, bevacqua2021guidelines, Rohrbeck2021, Sando2022}.
Consequently, joint risks may be underestimated unless tail \emph{dependence} is explicitly modelled \citep{Zhang2024}. 
The strength of tail dependence between two random vectors is commonly summarised by the coefficient of asymptotic dependence \(\chi\in[0,1]\) \citep[see e.g.,][]{Coles1999}.
For two random variables $X_1$ and $X_2$ with distribution functions $F_1$ and $F_2$, respectively, $\chi$ is defined as
\begin{equation} \label{eq:chi}
  \chi \;=\; \lim_{u\to 1} \chi(u) \;=\; \lim_{u\to 1}\Pr\left\{F_1(X_1)>u \mid F_2(X_2)>u\right\}.
\end{equation}
When $\chi > 0$, the variables $X_1$ and $X_2$ are \emph{asymptotically dependent} (AD), while $\chi=0$ corresponds to \emph{asymptotic independence} (AI) between $X_1$ and $X_2$.
Although \(\chi\) is a useful summary of pairwise extremal dependence, it does not fully characterise the multivariate tail dependence structure or support inference on joint exceedance probabilities, for which a full probabilistic model is required.

We consider a setting with $D$ random vectors, $\boldsymbol{X}_1, \ldots, \boldsymbol{X}_D$, of equal length $d$, but with potentially different joint tail behaviours. 
For instance, we may observe environmental variables (e.g.,\@ rainfall and wind speed) that exhibit non-stationary extremal dependence with respect to their sampling index or location: this could be time, or a fixed spatial location. 
Interest then lies in understanding the variation in tail behaviour across the $D$ vectors and, where feasible, to pool information to reduce estimation uncertainty. 
However, despite the importance of identifying groups of random vectors with similar extremal dependence behaviour, there are relatively few tools available for clustering multivariate extremes.
Existing approaches, such as that of \citet{Vignotto2021}, are often restricted to asymptotic dependence, low-dimensional settings, or non-parametric summaries of tail behaviour, making them less suitable for complex environmental applications where asymptotic independence may also be present. 
To address this, we leverage the multivariate conditional extremes (CE) framework of \citet{Heffernan2004}, which is a popular and flexible model for multivariate extremes, to propose a novel clustering algorithm for multivariate extremes.
To illustrate the efficacy of our approach, we consider a motivating application with variables observed at $D$ spatial locations (see Section~\ref{sec:application}), but our method is more broadly applicable to other fields, e.g.\ finance, where clustering multivariate extremes is of interest. 

\citet{Heffernan2004} described the tail behaviour of a random vector \emph{conditional} on one of its components being extreme, i.e.\ exceeding a suitably high threshold. 
This framework can flexibly capture both AD and AI, and is both conceptually simple and computationally efficient.
Spatial and spatio-temporal extensions have been developed to handle high-dimensional data \citep{Simpson2021, Wadsworth2022}, and adaptations that incorporate covariates, hierarchical structures, and mixtures have further increased the models' flexibility \citep{Richards2023, Tendijck2023, Talento2025}.
As a result, CE models have seen widespread application in settings where AI as well as AD may be prevalent, including river flow and flood risk \citep{Gilleland2013, Jane2020}, rainfall \citep{Richards2022}, sea-surface temperatures \citep{Simpson2023,Kakampakou2024}, heatwaves \citep{Winter2016}, neuroscience \citep{Guerrero2023, Talento2025}, and finance \citep{Hilal2014}. 
The broad applicability of CE models makes them well-suited to our motivating application.
However, the scarcity of joint extreme observations, particularly in high dimensions, and the number of model parameters, often results in high uncertainty in estimates of the CE model.

Clustering is a popular tool in the extremes literature, where data sparsity and low sample sizes provide a natural motivation, and we distinguish three types of extremal clustering methods.
\emph{Univariate} methods group data by marginal tail behaviour, often through summaries such as return levels or parameter estimates \citep{Coles2001}, and have been used in environmental and financial applications \citep{Alonso2014-rr, deCarvalho2023, Zheng2023-jf}.
\emph{Spatial} methods cluster entire spatial fields or partition the spatial domain via pairwise tail dissimilarities, such as the F-madogram \citep{Cooley2006variograms, Bernard2013}, or by fitting non-stationary extremal dependence models \citep{Carreau2017, Rohrbeck2021, Dupuis2023, Shao2025}. 
Finally, \emph{multivariate} methods cluster on joint tail summaries or fitted multivariate extremal dependence models. 
\citet{Vignotto2021} consider clusters of bivariate extremes of precipitation and wind speeds across spatial locations in Great Britain and Ireland.
Their nonparametric approach estimates pairwise Kullback--Leibler ($\KL{}$) \citep{Kullback1951} divergences for bivariate precipitation and wind extremes (see \citet{Zscheischler2021}) and then applies the $k$-medoids algorithm. 
However, their method is restricted to the bivariate case and, being non-parametric, lacks a generative model for tail dependence inference. 
Also, it is undefined where no joint extreme events are observed, which can occur frequently in environmental applications, where AI is prevalent.

We address the limitations of existing methods by developing a clustering framework for multivariate extremes that leverages the CE framework. 
This enables flexible, parametric exploration of tail dependence in arbitrary dimensions, in contrast to the existing non-parametric approach of \citet{Vignotto2021}, and overcomes the data sparsity issue often associated with drawing interpretations from fitted CE models.
Many classical models and clustering frameworks assume AD \citep{Vignotto2021,Boulin2025}, but recent literature argues that AD models are inappropriate for environmental data \citep{Opitz2016,Dawkins2018,Huser2024}. 
Our clustering method can flexibly handle both AD and AI. 

To compare fitted CE models, we propose a dissimilarity measure that is computationally tractable in high dimensions.
We employ the skew-geometric Jensen-Shannon divergence ($\JSGa{}$) of \citet{Nielsen2019}, which yields a closed-form divergence for multivariate Gaussian distributions.
Combined with the working Gaussianity assumption of \citet{Heffernan2004}, our proposed extremal skew-geometric $\JSGa{}$ divergence measure is computationally efficient, non-negative, and bounded, which makes it well suited to clustering of multivariate extremes using traditional similarity-based algorithms.
Our clustering approach delivers a unified and statistically principled tool for exploratory analysis of multivariate extremes, facilitating the identification of groups with similar tail dependence. 

The remainder of this paper is structured as follows.
Section~\ref{sec:method} introduces our novel dissimilarity measure for multivariate CE models, and we illustrate its use for clustering of multivariate extremes. 
Section~\ref{sec:sim} provides a simulation study where we demonstrate that our approach outperforms the clustering method of \citet{Vignotto2021}.
In Section~\ref{sec:application}, we apply our clustering framework to Irish meteorological data, and identify sites with similar tail dependence between precipitation and wind speeds.
Section~\ref{sec:discuss} concludes with a discussion of our findings, the limitations of our method, and potential avenues for future work. 

\section{Methodology}
\label{sec:method}

In this section, we introduce our clustering framework. 
Section~\ref{subsec:ce_model} provides background on the conditional extremes (CE) model of \citet{Heffernan2004} and \citet{Keef2013}.
Section~\ref{subsec:ce_setup} defines the setup for our clustering model and describes how inference is performed.
In Section~\ref{subsec:div}, we describe our novel dissimilarity measure for multivariate extremes, which uses the skew-geometric Jensen-Shannon divergence to efficiently quantify dissimilarity between fitted CE models.
Finally, in Section~\ref{subsec:clustering}, we outline how to perform clustering using this dissimilarity measure, and provide practical advice on choosing the number of clusters.

\subsection{Background to multivariate conditional extremes}
\label{subsec:ce_model}

Like many other models for extremal dependence \citep[see, e.g.,][]{Davison2015}, the conditional extremes framework of \citet{Heffernan2004} is defined for data on standardised margins, which ensures that differences in marginal scale do not affect estimation of the dependence structure.
As introduced in \citet{Keef2013}, a pragmatic choice of margins for CE models is standard Laplace, as it allows for modelling of both positive and negative dependence.
Let $\boldsymbol{X} \coloneqq (X_1, \ldots, X_d) \in \mathbb{R}^d$ be a $d$-dimensional random vector representing observed data with marginal distribution functions $F_{X_1}, \ldots, F_{X_d}$.
We apply the probability integral transformation to transform $\boldsymbol{X}$ to a random vector with standard Laplace margins, denoted by $\boldsymbol{Y} \coloneqq (Y_1, \ldots, Y_d) \in \mathbb{R}^d$.
Formally, for $i = 1, \ldots, d$, 
\begin{equation}\label{eq:laplace}
  Y_i \coloneqq \begin{cases}
    \log\left\{2F_{X_i}(X_i)\right\}, &\text{ for } X_i < F_{X_i}^{-1}(0.5), \\
    -\log\left[2\left\{ 1 - F_{X_i}(X_i)\right\} \right], &\text{ for } X_i \ge F_{X_i}^{-1}(0.5). \\
  \end{cases}
\end{equation}

Let $\boldsymbol{Y}_{-i} \coloneq \boldsymbol{Y} \backslash Y_i \in \mathbb{R}^{d-1}$ denote $\boldsymbol{Y}$ with its $i^{\text{th}}$ component removed, for $i \in \{1, \ldots, d\}$.
With all operations below taken component-wise, \citet{Heffernan2004} assumed that there exist parameter vectors $\boldsymbol{\alpha}_{\mid i} \coloneqq \left\{\alpha_{j \mid i} : j\in\{1,\dots,d\}\backslash i \right\} \in [-1, 1]^{d-1}$ and $\boldsymbol{\beta}_{\mid i} \coloneqq \left\{\beta_{j \mid i} : j\in\{1,\dots,d\}\backslash i \right\} \in (-\infty, 1]^{d-1}$ such that, for fixed $\boldsymbol{z} \in \mathbb{R}^{d-1}$,
\begin{equation}\label{eq:ce_limiting_dist}
  \mathbb{P}\!\left(
  \frac{\boldsymbol{Y}_{-i} - \boldsymbol{\alpha}_{\mid i} Y_{i}}
  {Y_{i}^{\boldsymbol{\beta}_{\mid i}}}
  \le \boldsymbol{z},\; 
  Y_i - u > y \;\Big|\; Y_i > u
  \right)
  \longrightarrow
  G_{\mid i}(\boldsymbol{z})\,\exp{(-y)}
  \quad\text{as } u \to \infty,
\end{equation}
where the distribution function $G_{\mid i}\left(\boldsymbol{z}\right)$ is non-degenerate on $\boldsymbol{z} \in \left[ -\infty, \infty \right)^{d-1}$.
Equation~\eqref{eq:ce_limiting_dist} implies that, in the limit as $u\to\infty$, the excesses $Y_i-u$ and the standardised residuals 
$\boldsymbol{Z}_{\mid i} \coloneqq Y_i^{-\boldsymbol{\beta}_{\mid i}}\left(\boldsymbol{Y}_{-i} - \boldsymbol{\alpha}_{\mid i} Y_{i}\right)$ are independent.

Assuming that the limit in Equation~\eqref{eq:ce_limiting_dist} holds exactly for a large threshold $u_i > 0$ gives the heteroscedastic regression model
\begin{align}\label{eq:ce_model}
\bigl(\boldsymbol{Y}_{-i} \mid Y_i = y \bigr) &= \boldsymbol{\alpha}_{\mid i}y + y^{\boldsymbol{\beta}_{\mid i}}\boldsymbol{Z}_{\mid i},\qquad \text{ for } y > u_i,
\end{align}
with residual vector $\boldsymbol{Z}_{\mid i} \sim G_{\mid i}$.
The parameter vector $\boldsymbol{\alpha}_{\mid i}$ describes the strength of extremal association between components of $\boldsymbol{Y}_{-i}$ and the conditioning variable $Y_i$, while $\boldsymbol{\beta}_{\mid i}$ determines the spread induced by large values of $Y_i$. 
Some special cases (for $j\in\{1,\ldots,d\} \backslash i$) of the CE model are:
$\alpha_{j\mid i}=0$ and $\beta_{j\mid i}=0$, which imply asymptotic independence between $Y_j$ and $Y_i$, corresponding to $\chi=0$;
$\alpha_{j\mid i}=1\ ( -1)$ and $\beta_{j\mid i}=0$, which corresponds to positive (negative) asymptotic dependence, and
$-1<\alpha_{j\mid i}<1$, which suggests asymptotic independence between $Y_j$ and $Y_i$, with the strength of extremal dependence weakening with $\alpha_{j \mid i}$. 

\subsection{Framework and model inference}
\label{subsec:ce_setup}

In our motivating spatial application, we consider a CE model that varies across sites and across conditioning variables.
To denote spatial variation, we include the additional subscript $s \in \{1, \ldots, D\}$, where $D$ is the number of spatial locations, so that $Y_{i,s}$ denotes the variable $Y_i$ observed at site $s$.
This leads to a spatial extension of Equation~\eqref{eq:ce_model}, with site-specific random vector $\boldsymbol{Y}_{s} \coloneq \left(Y_{1,s}, \ldots, Y_{d,s}\right) \in \mathbb{R}^d$, and, for $i = 1, \ldots, d$,
\begin{equation} \label{eq:ce_model_spat}
  \bigl(\boldsymbol{Y}_{-i, s} \mid Y_{i, s} = y\bigr)
    = \boldsymbol{\alpha}_{\mid i, s} \, y
      + y^{\boldsymbol{\beta}_{\mid i, s}} \, \boldsymbol{Z}_{\mid i, s},
      \quad y > u_{i, s},
\end{equation}
where $\boldsymbol{Y}_{-i, s} \coloneqq \boldsymbol{Y}_{s} \backslash Y_{i, s}$, and $\boldsymbol{\alpha}_{\mid i, s}$, $\boldsymbol{\beta}_{\mid i, s}$, $\boldsymbol{Z}_{\mid i, s} \sim G_{\mid i, s}$, and $u_{i, s}$ are the site-specific analogues of the regression parameters, residuals, and exceedance threshold in Equations~\eqref{eq:ce_limiting_dist} and~\eqref{eq:ce_model}.
All operations involving vectors in Equation~\eqref{eq:ce_model_spat} are taken componentwise.
Note that we do not assume spatial smoothness in this regression model.

As a working assumption for tractable estimation of $\boldsymbol{\alpha}_{\mid i, s}$ and $\boldsymbol{\beta}_{\mid i, s}$, we follow \citet{Heffernan2004} and assume that the residual vector $\boldsymbol{Z}_{\mid i, s}$ in Equation~\eqref{eq:ce_model_spat} is multivariate Gaussian with mean vector $\boldsymbol{\mu}_{\mid i, s} \in \mathbb{R}^{d-1}$ and covariance matrix $\Sigma_{\mid i, s} \in\mathbb{R}^{(d-1)\times(d-1)}$ for $i = 1, \ldots, d$ and $s = 1, \ldots, D$.
This assumption is standard in the conditional extremes literature and underpins a wide range of applications and extensions \citep[e.g.,][]{Keef2012_flooding, Keef2013, Gilleland2013, Hilal2014, Winter2016, Tendijck2023}.
Under this working model, likelihood-based estimates of the CE parameters are consistent, even if the Gaussianity assumption is not strictly satisfied.

This working assumption is the key to deriving a closed-form divergence measure for computationally efficient clustering; see Section~\ref{subsec:div}.
It follows that 
\begin{equation}\label{eq:ce_norm}
  \bigl(\boldsymbol{Y}_{-i,s}\mid Y_{i,s}=y\bigr)
  \sim
  \mbox{MVN}_{d-1}
  \left(
    \boldsymbol{\alpha}_{\mid i,s}y
    +
    y^{\boldsymbol{\beta}_{\mid i,s}}
    \boldsymbol{\mu}_{\mid i,s},
    \;
    \Omega_s(y)
  \right),
  \quad y>u_{i,s},
\end{equation}
where
$\Omega_s(y)
=
\operatorname{diag}\!\left(
y^{\boldsymbol{\beta}_{\mid i,s}}
\right)
\Sigma_{\mid i,s}
\operatorname{diag}\!\left(
y^{\boldsymbol{\beta}_{\mid i,s}}
\right)$,
and $\boldsymbol{\mu}_{\mid i,s}$ and $\Sigma_{\mid i,s}$ are, respectively, the site-specific residual mean vector and covariance matrix for conditioning variable $i$, $i=1,\ldots,d$, at site $s$, $s=1,\ldots,D$.
We emphasise that, in practice, Gaussianity of the residuals is not required to hold exactly. 
The CE model is asymptotic in nature, and the assumed residual representation should therefore be viewed as an approximation at finite thresholds, with the quality of the approximation expected to improve as the threshold increases \citep{Heffernan2004, Keef2013}.
The good empirical performance observed in our simulation study (Section~\ref{sec:sim}) and application (Section~\ref{sec:application}) suggests that our proposed clustering method is robust to misspecification of residual Gaussianity.
Moreover, after estimation of the CE parameters, the residual distribution $G_{\mid i,s}$ is treated non-parametrically using the empirical joint distribution of the observations
\[
\boldsymbol{Z}_{\mid i, s} = \frac{\boldsymbol{Y}_{-i, s} - \boldsymbol{\hat{\alpha}}_{\mid i, s} Y_{i, s}}{Y_{i, s}^{\boldsymbol{\hat{\beta}}_{\mid i, s}}},
\]
for $Y_{i, s} > u_{i, s}$, as in \citet{Keef2013}. 

To compare tail dependence across different random vectors $\boldsymbol{Y}_{s}$ at different sites $s$, we compare the conditional distributions in Equation~\eqref{eq:ce_norm}, i.e.,\@ the assumed multivariate Gaussian distributions parameterised by $\boldsymbol{\alpha}_{\mid i, s},\boldsymbol{\beta}_{\mid i, s},\boldsymbol{\mu}_{\mid i, s}$ and $\Sigma_{\mid i, s}$, for $i = 1, \ldots, d$. 

We estimate parameters $\boldsymbol{\alpha}_{\mid i, s}, \boldsymbol{\beta}_{\mid i, s}, \boldsymbol{\mu}_{\mid i, s}, \Sigma_{\mid i, s}$, for each conditioning variable $i = 1, \ldots, d$ and site $s = 1, \ldots, D$, using maximum likelihood estimation.
In practice, we set the threshold $u_{i, s} \coloneqq u_i$ to the $q^{\text{th}}$ standard Laplace quantile for all $s$, as this is required to build a symmetric divergence measure for clustering (see Section~\ref{subsec:div}).
Choosing $q$ involves the classic bias-variance tradeoff: $q$ must be sufficiently low as to leave enough excesses above $u_i$ for reliable estimation, but high enough to reduce bias induced when the asymptotic arguments in Equation~\eqref{eq:ce_limiting_dist} are not satisfied. 
To pick $q$, we use threshold stability plots and assess the stability of parameter estimates at different thresholds \citep{Southworth2012}. 
Using the bootstrap procedure of \citet{Heffernan2004}, we obtain standard errors for parameter estimates at each site and choose the highest value of $q$ such that the estimates are stable and the standard errors are reasonably small across sites. 

\subsection{A divergence measure for multivariate conditional extremes}
\label{subsec:div}

There are several qualities we desire from a divergence measure based on the conditional distribution in Equation~\eqref{eq:ce_norm}:
(i) the measure should be non-negative and (ii) symmetric so that it can be interpreted as a distance \citep{Vijendra2015};
(iii) the measure should have an upper bound, as large divergences can lead to numerical instability during clustering \citep{Thiagarajan2025};
(iv) it should be cheap to compute, which we provide through a closed-form solution for our divergence. 

\citet{Nielsen2019} introduced the skew-geometric Jensen-Shannon divergence ($\JSGa{}$), which is a generalisation of the Jensen-Shannon ($\JS{}$) divergence \citep{Lin1991}. 
Specifically, we use the dual form of the $\JSGa{}$ divergence \citep{Deasy2020}, which is defined between two continuous $(d-1)$-variate distribution functions, say $H$ and $H^*$, as
\begin{equation}\label{eq:jsga}
  \JSGa{(H \Vert H^*)} = (1- \lambda) \KL{\left\{G_{\lambda}(H, H^*) \Vert H\right\}} + \lambda \KL{\left\{G_{\lambda}(H, H^*) \Vert H^*\right\}},
\end{equation}
where $G_{\lambda}(H, H^*)$ denotes the distribution whose density is proportional to the weighted geometric mean $h^{1-\lambda}(h^*)^{\lambda}$ of the densities $h$ and $h^*$ corresponding to $H$ and $H^*$, respectively, with $\lambda \in [0, 1]$.
The Kullback-Leibler ($\KL{}$) divergence between $H$ and $H^*$ is given by
\begin{equation}\label{eq:kl}
  \KL{(H \Vert H^*)} = \int_{\mathbb{R}^{d-1}} h(\boldsymbol{y}) \log\left(\frac{h(\boldsymbol{y})}{h^*(\boldsymbol{y})}\right) \mathrm{d}\boldsymbol{y}.
\end{equation}
In our context, we replace $H$ and $H^*$ by the conditional multivariate Gaussian distributions induced by the CE framework at two different sites, say $s$ and $s^*$; see Equation~\eqref{eq:ce_norm}.

The weight $\lambda$ skews the geometric mean $G_{\lambda}(H,H^*)$ towards $H$ as $\lambda \to 0$ (and towards $H^*$ as $\lambda \to 1$).
One possible advantage of specifying $\lambda$ is that we can design our divergence measure to place more weight at sites $s$ with more reliable CE model fits, as determined by, for example, the effective sample size.
In this paper, we set $\lambda = 0.5$ throughout, as all sites have the same number of observations in both our simulations and case study.

For two multivariate Gaussian distributions, the $\KL{}$ divergence in Equation~\eqref{eq:kl} has a closed-form solution \citep{Duchi2007}.
The $\JSGa{}$ divergence exploits the property that the (weighted) product of two distributions from the same exponential family is proportional to another member of that family, and hence, after normalisation, is itself an exponential-family distribution \citep{Nielsen2019}.
In particular, when $H$ and $H^*$ are multivariate Gaussian distributions, the corresponding density of the mixture $G_{\lambda}$ is Gaussian after normalisation, and so the $\KL{}$ divergences in Equation~\eqref{eq:jsga} are available in closed-form.
For two multivariate Gaussian distributions $H \sim \mbox{MVN}(\boldsymbol{m}, \Omega)$ and $H^* \sim \mbox{MVN}(\boldsymbol{m}^*, \Omega^*)$ in $\mathbb{R}^d$ with mean vectors $\boldsymbol{m}, \boldsymbol{m}^* \in \mathbb{R}^d$ and covariance matrices $\Omega, \Omega^* \in \mathbb{R}^{d \times d}$, \citet{Nielsen2019} showed that the $\JSGa{}$ divergence introduced in Equation~\eqref{eq:jsga} has the closed-form solution 
\begin{align*}
\JSGa{}(H \Vert H^*)
  &= \frac{1}{2}\Biggl\{
      (1-\lambda)\boldsymbol{m}^{\mathrm T}\Omega^{-1}\boldsymbol{m}
      + \lambda(\boldsymbol{m}^*)^{\mathrm T}(\Omega^*)^{-1}\boldsymbol{m}^*
      \notag\\
  &\qquad
      - \boldsymbol{m}_{\lambda}^{\mathrm T}\Omega_{\lambda}^{-1}\boldsymbol{m}_{\lambda}
      + \log\left(
          \frac{\abs{\Omega}^{1-\lambda}\abs{\Omega^*}^{\lambda}}
               {\abs{\Omega_{\lambda}}}
        \right)
      \Biggr\}, \\
\boldsymbol{m}_{\lambda}
  &= \Omega_{\lambda}
     \left\{
       (1-\lambda)\Omega^{-1}\boldsymbol{m}
       + \lambda(\Omega^*)^{-1}\boldsymbol{m}^*
     \right\}, \text{ and} \\
\Omega_{\lambda}
  &= \left\{
       (1-\lambda)\Omega^{-1}
       + \lambda(\Omega^*)^{-1}
     \right\}^{-1}.
\numberthis \label{eq:js_norm_mvn1}
\end{align*}

Moreover, as a divergence measure, $\JSGa{(H \Vert H^*)} \ge 0$, with equality if and only if $H = H^*$.
Finally, the $\JSGa{}$ is bounded above by $\JSGa{(H \Vert H^*)} \leq \frac{1}{2}\max\{\lambda,\,1-\lambda\}\mathrm{J}(H \Vert H^*)$, where the Jeffreys' divergence $\mathrm{J}(H \Vert H^*) = \KL{(H \Vert H^*)} + \KL{(H^* \Vert H)}$ is simply the sum of the forward and backward $\KL{}$ divergences \citep{Nielsen2025}.

Now, suppose that for a specified conditioning variable $i \in \{1, \ldots, d\}$, we have fitted the CE model at two different sites, $s$ and $s^*$, and we want to compare their induced extremal dependence (i.e.\ the joint tails of $\boldsymbol{Y}_{i, s}$ and $\boldsymbol{Y}_{i, s^*}$; see Equation~\eqref{eq:ce_norm}). 
Using the $\JSGa{}$ divergence in Equation~\eqref{eq:js_norm_mvn1}, we define the \textit{conditional extremal skew-geometric Jensen-Shannon divergence} ($\cJSGa{s, s^*}$) between two sites $s$ and $s^*$, conditional on the $i$-th variable being extreme, as
\begin{equation}\label{eq:cjsga}
  \cJSGa{s, s^*}(y) = \JSGa{}\left(H_{i, s}(y) \Vert H_{i, s^*}(y)\right),\qquad y > u_i,
\end{equation}
where $H_{i, s}(y)$ and $H_{i, s^*}(y)$ are the conditional $(d-1)$-variate Gaussian distributions at sites $s$ and $s^*$, respectively; see Equation~\eqref{eq:ce_norm}.
Additionally, $\cJSGa{s, s^*}(y) \ge 0$ for all $y > u_i$, with equality if and only if the corresponding conditional distributions coincide.
Equation~\eqref{eq:cjsga} has a closed-form expression.
From Equation~\eqref{eq:ce_norm}, this is obtained by substituting into Equation~\eqref{eq:js_norm_mvn1} the conditional mean vectors
$\boldsymbol{m}_{s}(y)=\mvmean{s}$ and
$\boldsymbol{m}_{s^*}(y)=\mvmean{s^*}$,
and the conditional covariance matrices
$\Omega_s(y)$ and $\Omega_{s^*}(y)$.
Specifically,
\begin{align*}
\Omega_{\lambda}(y)
  &= \left[
       (1-\lambda)\,\Omega_s(y)^{-1}
       +\lambda\,\Omega_{s^*}(y)^{-1}
     \right]^{-1},\\[6pt]
\boldsymbol{m}_{\lambda}(y)
  &= \Omega_{\lambda}(y)
     \left[
       (1-\lambda)\,\Omega_s(y)^{-1}\boldsymbol{m}_{s}(y)
       +\lambda\,\Omega_{s^*}(y)^{-1}\boldsymbol{m}_{s^*}(y)
     \right].
\end{align*}

The $\cJSGa{s, s^*}(y)$ depends on $y > u_i$, the value of the corresponding conditioning variable. 
Therefore, to obtain a single measure of dissimilarity between the CE models at the two sites $s$ and $s^*$ for the $i$-th conditioning variable, we define the expected extremal divergence
\begin{align*}
  \eJSGa{s, s^*} &= \mathbb{E}\left(\cJSGa{s, s^*}(Y_{i,s}) \mid Y_{i,s} > u_i\right) \\
                 &= \mathbb{E}\left(\cJSGa{s^*, s}(Y_{i,s^*}) \mid Y_{i,s^*} > u_i\right) \\
                 &= \int_{u_i}^{\infty} \cJSGa{s, s^*}(y) h(y \mid Y_{i,s} > u_i) \mathrm{d}y, \numberthis \label{eq:jsga_integral}
\end{align*}
where $h(y \mid Y_{i, s} > u_i)$ is the density of $Y_{i, s} \mid (Y_{i, s} > u_i)$, and similarly $h(y \mid Y_{i, s^*} > u_i)$ denotes the density of $Y_{i, s^*} \mid (Y_{i, s^*} > u_i)$.
With standardised Laplace margins, this conditional distribution does not depend on the site $s$, and hence $h(y \mid Y_{i, s} > u_i) = h(y \mid Y_{i, s^*} > u_i)$ for all $s, s^*$, ensuring symmetry of our dissimilarity measure.
The proposed expected divergence inherits several desirable properties from the conditional divergence. 
Since $\cJSGa{s,s^*}(y)\ge0$ for all $y>u_i$, it follows immediately that $\eJSGa{s,s^*} \ge 0$.
Boundedness is established in the following proposition.
\begin{proposition} \label{prop:bounded}
Assume that the conditional covariance matrices in Equation~\eqref{eq:ce_norm}, $\Sigma_{\mid i,s}$ and $\Sigma_{\mid i,s^*}$, are positive definite. 
Then, for any pair of sites $(s,s^*)$, the expected extremal skew-geometric Jensen-Shannon divergence is bounded, i.e.,
\[
0\le \eJSGa{s,s^*}<\infty.
\]
\end{proposition}
Proposition~\ref{prop:bounded} establishes that the expected divergence is well-defined under mild regularity conditions. 
Its proof is given in Appendix~\ref{sec:proof_boundedness} of the Supplementary Material.

We estimate the integral in Equation~\eqref{eq:jsga_integral} using Monte Carlo methods, and now discuss its computational cost.
Sampling from $h(y \mid Y_{i,s} > u_i)$ has constant cost, as this distribution is exponential under the Laplace marginal assumption. 
The dominant computational cost arises from evaluating $\cJSGa{s, s^*}(y)$ at each Monte Carlo draw via the closed-form expression in Equation~\eqref{eq:js_norm_mvn1}, which requires inversion or factorisation of $(d-1)\times(d-1)$ covariance matrices and hence has complexity $O\!\left((d-1)^3\right)$. 
Therefore, using $N_{\mathrm{MC}}$ Monte Carlo samples yields an overall computational cost of $O\!\left(N_{\mathrm{MC}}(d-1)^3\right)$ for a fixed conditioning variable $i$ and pair of sites $(s,s^*)$. 
In practice, very large values of $Y_{i, s}$ can lead to unstable estimates, and we want to avoid extrapolating beyond the range of the observations of $Y_{i, s}$ used to fit the CE model. 
Hence, for estimation, we truncate the limits in integral~\eqref{eq:jsga_integral}, so that the upper limit is the 99th percentile of the empirical distribution of $Y_{i, s}$ pooled across all sites. 

We evaluate the pairwise $\eJSGa{s, s^*}$ for all pairs of sites $(s, s^*)$ and for each conditioning variable $i \in \{1, \ldots, d\}$. 
These are then collected in a dissimilarity matrix:
\begin{equation} \label{eq:distance_matrix}
  M^{(i)} \coloneq \left(\eJSGa{s, s^*} \right)_{s, s^* = \{1, \ldots, D\}},
\end{equation}
where $D$ is the total number of sites, and $i = 1, \ldots, d$.

If desired, we can aggregate the $d$ dissimilarity matrices $M^{(1)}, \ldots, M^{(d)}$ into a single dissimilarity matrix $M$ by taking, for example, their element-wise average, i.e. 
\begin{equation} \label{eq:aggregate_matrix}
  M = \frac{1}{d} \sum_{i=1}^d M^{(i)}.
\end{equation}
This aggregated matrix $M$, or the individual matrices $M^{(i)}$, can then be used as input to a clustering algorithm, which we describe in Section~\ref{subsec:clustering}.
Aggregating allows us to pool information across multiple comparisons and forms of information, and is supported in the extremal clustering literature \citep{Mornet2017-is}.

\subsection{Clustering of multivariate extremes}
\label{subsec:clustering}

We describe the application of the Partitioning Around Medoids (PAM) algorithm to a generic dissimilarity matrix $M$, which may correspond either to $M^{(i)}$ for a fixed conditioning variable $i$, or to the aggregated matrix $M = \frac{1}{d}\sum_{i=1}^d M^{(i)}$, as defined in Equations~\eqref{eq:distance_matrix} and \eqref{eq:aggregate_matrix}, respectively.
As in \citet{Bernard2013} and \citet{Vignotto2021}, we use the PAM algorithm, which is a variant of the popular k-medoids algorithm of \citet{Kaufman1987}.
For cluster label $c = 1, \ldots, k$, $\kappa_c \in \{1, \ldots, D\}$ denotes its medoid index, and $\mathcal{C}_c \subseteq \{1, \ldots, D\}$ the set of site indices assigned to cluster $c$; the medoid site is the location with index $\kappa_c$.

For the dissimilarity matrix $M$, a brief description of the algorithm is as follows:
\begin{algorithm}[htb]
  \caption{Partitioning Around Medoids (PAM) algorithm applied to a dissimilarity matrix $M$}.
  \label{alg:kmedoids_kappa}
  \begin{algorithmic}[1]
    \State \textbf{Input:} dissimilarity matrix $M\in\mathbb{R}^{D\times D}$, number of clusters $k$.
    \State \textbf{Output:} Cluster assignments $\mathcal{C}_1,\dots,\mathcal{C}_k$, medoid indices $\boldsymbol{\kappa}=(\kappa_1,\dots,\kappa_k)$.
    \State \textbf{Initialise:} choose $k$ distinct medoid indices $\{\kappa_1,\dots,\kappa_k\}\subseteq\{1,\dots,D\}$ uniformly at random.
    \Repeat

    \State \textbf{Assignment:}
    \State Assign each site $s\in\{1,\dots,D\}$ to its closest medoid, i.e.\ assign $s$ to $\mathcal{C}_c$, where $c = \argmin_{\ell\in\{1,\dots,k\}} M_{s,\kappa_\ell}$.
      \For{$c=1,\dots,k$}
        \State \textbf{Update:}
        \State Update medoid index to minimise intra-cluster dissimilarity, i.e.\ $\kappa_c = \argmin_{p\in \mathcal{C}_c}\sum_{q\in \mathcal{C}_c} M_{p,q}$.
        \State \Comment{Break ties by choosing the smallest index.}
      \EndFor
    \Until{no observation changes cluster assignment.}
    \State \Return $\mathcal{C}_1,\dots,\mathcal{C}_k,\ \boldsymbol{\kappa}$.
  \end{algorithmic}
\end{algorithm}

The PAM algorithm takes a single hyperparameter, namely the number of clusters, $k$.
We follow the classical elbow criterion of \citet{Thorndike1953-ie}, based on the total within-cluster dissimilarity (TWD), to select $k$.
We find in our simulation study (Section~\ref{sec:sim}) that this method correctly identifies $k$.
Specifically, for a given dissimilarity matrix $M$, we define the $\mathrm{TWD}$ as
\begin{equation}\label{eq:twd}
  \mathrm{TWD}(k)
  =
  \sum_{c=1}^{k}
  \sum_{s \in \mathcal{C}_c}
  M_{s,\kappa_c},
\end{equation}
where $\kappa_c$ is the medoid index of cluster $\mathcal{C}_c$.
When $M=M^{(i)}$, this criterion is specific to conditioning variable $i$, whereas when M is an aggregated dissimilarity matrix (e.g.\ $M=d^{-1}\sum_{i=1}^{d}M^{(i)}$), it is based on the aggregated dissimilarities.
We plot $\mathrm{TWD}(k)$ against $k$ and select the optimal number of clusters as the value beyond which additional clusters yield only negligible reductions in $\mathrm{TWD}$, corresponding to the ``elbow'' of the curve \citep{Hastie2009}.

\section{Simulation study}
\label{sec:sim}

We illustrate the performance of our multivariate extremal clustering approach across multiple simulation studies. 
Clustering accuracy is evaluated using the adjusted Rand index (ARI) of \citet{Hubert1985-ys}. 
Unreported experiments showed that the $\mathrm{TWD}$ method correctly identified $k$, and so we proceed with $k$ fixed to the true number of clusters.
In Section~\ref{subsec:sim_gauss}, we generate data from a Gaussian copula for which the parameters of the CE model are known \citep{Keef2013}.
This allows us to verify that our clustering approach reduces estimation uncertainty and to investigate the sensitivity of the results to the choice of conditioning threshold.
Section~\ref{subsec:sim_mixture} extends the study to mixtures of Gaussian and $t$-copulas.
For the bivariate case, we compare our clustering approach to that of \citet{Vignotto2021}, while we also show how our method can be applied in higher dimensions and for more complex dependence structures.

\subsection{Gaussian copula setting}
\label{subsec:sim_gauss}

We consider a setting with $d=2$ variables observed at $D=12$ sites. 
Data are generated independently across sites from a Gaussian copula. 
For sites $s = {1, \ldots, 12}$, we sample 1000 observations of $(X_{1,s}, X_{2,s})$ from a bivariate Gaussian distribution with mean zero and correlation matrix whose off-diagonal elements $\rho^s_{\text{Gauss}} \in [0, 1]$ serve as site-specific correlation parameters.
We then apply the transformation in Equation~\eqref{eq:laplace}, with $F_X(\cdot)$ being the standard normal distribution function, to obtain $(Y_{1,s}, Y_{2,s})$ with standard Laplace margins.

The sites are assigned to one of three latent clusters of equal size, based on $\rho^s_{\text{Gauss}}$.
As a proof of concept, we choose well-separated values to ensure clear clustering.
Specifically, the first cluster, comprising sites 1 to 4, has correlation $\rho_{\text{Gauss}}^s=0.1, s = 1, \ldots, 4$; the second cluster has $\rho_{\text{Gauss}}^s=0.5$, for $s = 5, \ldots, 8$; and the third cluster has $\rho_{\text{Gauss}}^s=0.9$, for $s = 9, \ldots, 12$.
\citet{Keef2013} showed that the corresponding parameters of the CE framework are $\alpha_{\mid i} = (\rho_{\text{Gauss}}^s)^2$ and $\beta_{\mid i} = 1/2$, $i=1,2$, implying asymptotic independence between $Y_{1,s}$ and $Y_{2,s}$.

At each site $s = 1, \ldots, 12$, we estimate the CE model in Equation~\eqref{eq:ce_model}, with conditioning threshold $u_{i}$ as the standard Laplace $q$-quantile for \(q=0.9\) and \(q=0.99\).
Pairwise dissimilarities between each site are computed using the expected divergence in Equation~\eqref{eq:jsga_integral}, and we average over both conditioning directions, $i = 1, 2$, to obtain a single dissimilarity measure (see $M$ in Equation~\eqref{eq:aggregate_matrix}).
Clusters are identified using Algorithm~\ref{alg:kmedoids_kappa}.
We repeat this procedure 500 times, with performance assessed by comparing finite-sample CE estimates to the asymptotic truths. 

\begin{figure}[htb]
    \centering
    \includegraphics[width = 0.8\linewidth]{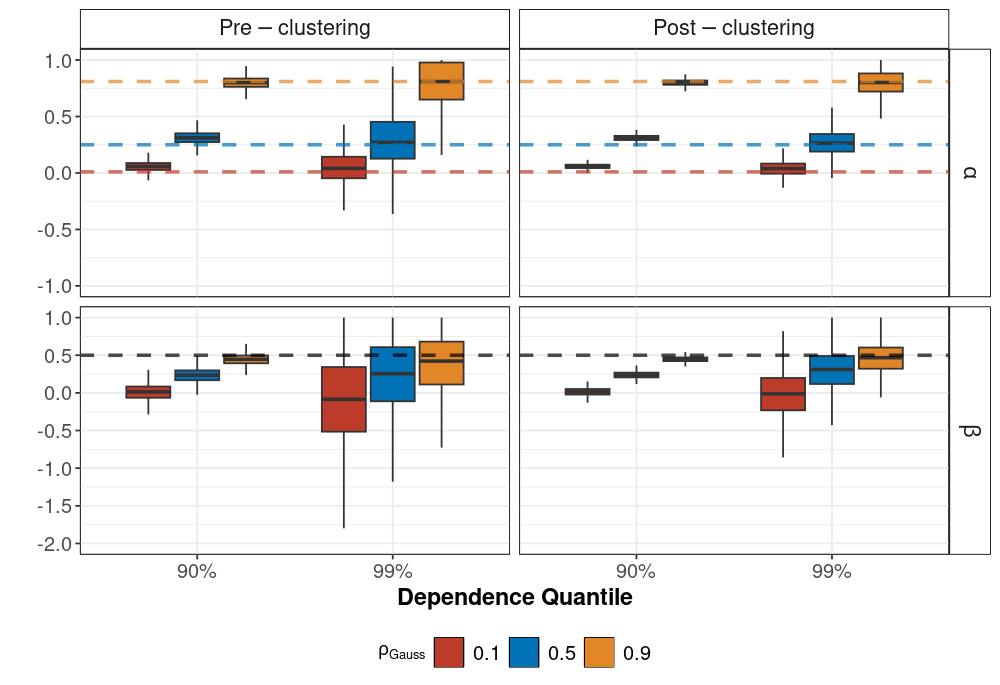}
    \caption{ 
    Boxplots of estimates of $\alpha_{\mid i}$ (top) and $\beta_{\mid i}$ (bottom) for the conditional extremes model pre- and post-clustering with bivariate Gaussian data simulated at 12 sites, belonging equally to three known clusters. 
    The x-axis gives $q$, the level of the conditioning quantile used to fit the model, and the boxes are coloured by the true Gaussian copula correlation parameter, $\rho_{\text{Gauss}}^s$.
    The horizontal lines indicate the true theoretical values of $\alpha_{\mid i}$ and $\beta_{\mid i}$, with the colours for $\alpha_{\mid i}$ corresponding with the true $\rho_{\text{Gauss}}^s$ values.
    }
   \label{fig:00_gauss_cop}
\end{figure}

Figure~\ref{fig:00_gauss_cop} shows boxplots of the estimated model parameters stratified by the true values of $\rho_{\text{Gauss}}^s$.
The results are shown pre- and post-clustering, with the former representing estimates obtained separately at each site, displayed according to the true cluster structure (i.e.\ the values of $\rho_{\text{Gauss}}^s$), and the latter derived by pooling data across all sites according to their estimated clustering (which was always correct in this setting).
By pooling data across sites, we reduce estimation uncertainty for both $\alpha_{\mid i}$ and $\beta_{\mid i}$.
Bias, as defined as the difference between the finite-sample estimate and the true theoretical parameter, decreases with increasing $\rho_{\text{Gauss}}^s$, indicating faster convergence to the asymptotic model under stronger dependence. 
We observe the expected bias-variance tradeoff in the choice of quantile for the CE model, with the higher quantile ($q=0.99$) yielding less bias, but with higher variance.

To assess the validity of the Gaussian residual assumption underlying the closed-form extremal divergence, we examine the fitted CE residuals using normal quantile-quantile (QQ) plots (see Figure~\ref{fig:diag_gauss} in the Supplementary Material).
These diagnostics suggest that the Gaussian approximation is reasonable at the threshold considered.
Moreover, the approximation improves with increasing values of $\rho_{\text{Gauss}}^s$, with residuals more closely following the theoretical normal quantiles under stronger dependence.
Importantly, clustering performance remains strong even where the Gaussian approximation is less suitable, indicating that the proposed extremal clustering method is robust to misspecification of the residual distribution.

\subsection{Mixture model setting}
\label{subsec:sim_mixture}

We extend our simulation design to mixtures of Gaussian and Student $t$-copulas \citep{Demarta2007-ap}, which induce asymptotic independence and asymptotic dependence, respectively, and whose dependence strength is determined by their respective correlation parameters $\rho_{\text{Gauss}}^s$ and $\rho_{\text{t}}^s \in [0, 1]$.
This construction provides a more realistic and challenging extremal dependence setting than the Gaussian copula example in Section~\ref{subsec:sim_gauss}, while still remaining within the conditional extremes (CEV) framework; see, for example, \citet{Tendijck2023} for a discussion of mixture structures in CE modelling.
Importantly, the mixture setting encompasses both asymptotic dependence (AD) and asymptotic independence (AI).
In particular, the model is asymptotically independent only in the special case where $\rho_{\mathrm{t}_1} = \rho_{\mathrm{t}_2} = 0$, whereas otherwise the extremal behaviour is dominated by the asymptotically dependent $t$-copula component.
Consequently, this simulation setting is generally favourable to the method of \citet{Vignotto2021}, which is specifically designed for AD settings.
Nevertheless, as shown below, our proposed method consistently outperforms the competing approach across the considered scenarios.

To obtain a sample of size $n = 1000$ from the mixture copula, at each site $s =1, \ldots, 12$, we combine $n/2 = 500$ samples from the Gaussian copula described in Section~\ref{subsec:sim_gauss} (with correlation parameter $\rhou{Gauss}^s$), and $n/2 = 500$ observations from a Student $t$-copula with 3 degrees of freedom and correlation $\rhou{t}^s$, before applying the Laplace transformation in Equation~\eqref{eq:laplace} to obtain the pairs $(Y_{1,s}, Y_{2,s})$ at site $s$.

In Section~\ref{subsubsec:sim_competing_methods}, we compare our method to the leading extremal competitor in the bivariate case \citep{Vignotto2021} 
Finally, Section~\ref{subsubsec:sim_realistic} considers a setting with $D = 60$ locations across three unequally sized clusters, and locations in the same cluster exhibiting perturbations in their correlation parameters.
Performance is evaluated using the ARI, with 500 replicated experiments per setting.

\subsubsection{Comparison to \citet{Vignotto2021}}
\label{subsubsec:sim_competing_methods}

When $d = 2$, we compare our extremal clustering method to that of \citet{Vignotto2021}.
We generate $n = 1000$ observations at $D = 12$ sites, using the mixture of Gaussian and t-copulas (with Laplace margins), as described above. 
Two clusters of sites $s = {1, \ldots, 6}$ and $s = {7, \ldots, 12}$ are defined with different values of $\rhou{t}^s$. 
We set $\rhou{t}^s = \rho_{\text{t}_1}$ for $s = 1, \ldots, 6$ and $\rhou{t}^s = \rho_{\text{t}_2}$ for $s = 7, \ldots, 12$, while $\rhou{Gauss}^s$ is the same for both clusters. 
We consider a range of extremal dependence scenarios defined over a grid of values $\left(\rhou{Gauss}^s \in \{0.1, 0.2, \ldots, 1.0\}, \rho_{\text{t}_1}^s \in \{0.1, 0.3, \ldots, 0.9\}, \rho_{\text{t}_2}^s \in \{0, 0.2, \ldots, 0.8\}\right)$.
We estimate the CE model using $q = 0.9$, and clusters were obtained using the aggregated dissimilarity matrix $M$ in Equation~\eqref{eq:aggregate_matrix}.

\begin{figure}[htb]
    \centering
    \includegraphics[width = 0.8\linewidth]{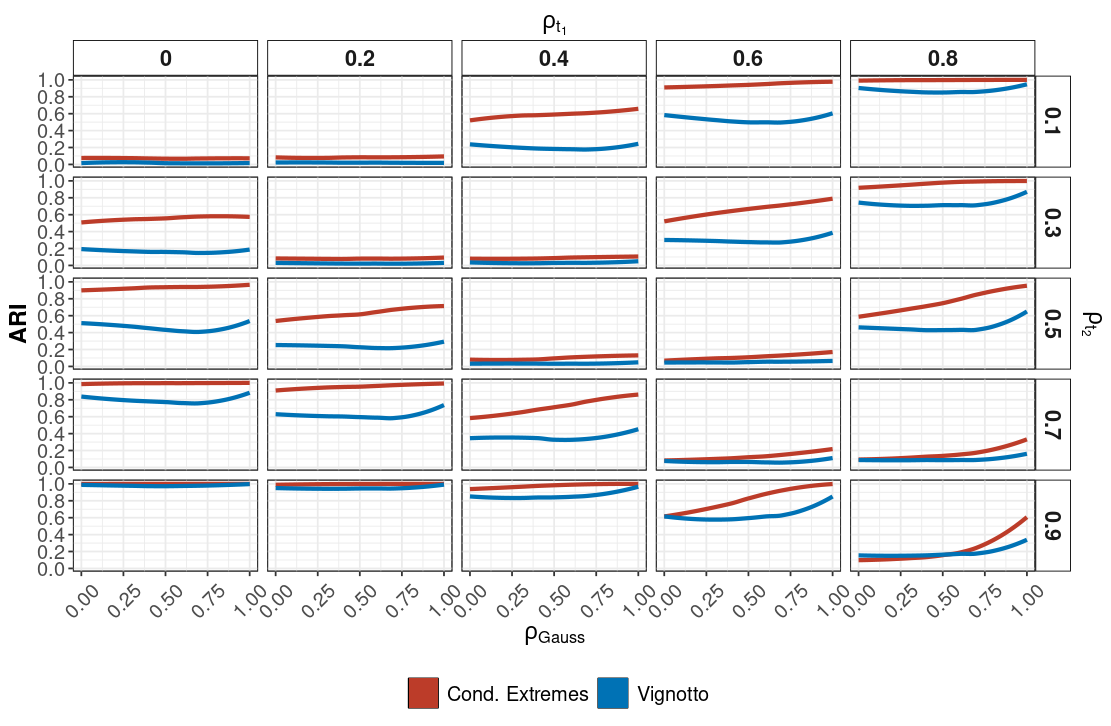}
    \caption{
    Comparison of our conditional extremes-based approach and the method of \citet{Vignotto2021}.
    The x-axis represents the Gaussian correlation parameter $\rhou{Gauss}$, with facet labels indicating the t-copula correlation parameters, $\rho_{\text{t}_1}$ and $\rho_{\text{t}_2}$, for the two true clusters.
    The smoothed lines show the average ARI across bootstrap samples, for both methods, with smoothing performed using a generalised additive model.
    }
    \label{fig:01_ce_vs_vi}
\end{figure}

The results are shown in Figure~\ref{fig:01_ce_vs_vi}. 
Across the range of considered bivariate extremal dependence structures, our method consistently outperforms that of \citet{Vignotto2021}. 
Both methods have high ARI when the cluster-wise difference in $\rhou{t}^s$ is larger, since greater disparity naturally makes clusters more distinct and clustering easier. 
Both methods perform better when the Gaussian and t-copula correlations are higher.
This may be because the Gaussian copula becomes more similar to the t-copula for higher values of $\rhou{Gauss}$, thus allowing the clustering approach to focus on differences in the t-copula.

To further illustrate the flexibility of the proposed framework beyond the bivariate setting, we additionally consider a three-dimensional simulation study with more general correlation structures (see Section~\ref{sec:sim_extension} in the Supplementary Material).

\subsubsection{Extension to more sites}
\label{subsubsec:sim_realistic}

We now design another simulation study to better mimic the structure of the Irish meteorological dataset, to be analysed in Section~\ref{sec:application}. 
Data are generated at $D = 60$ sites, each with $n = 1000$ observations of $d = 2$ variables, using the mixture of Gaussian and t-copulas.
The 60 sites are partitioned into three unequally-sized clusters of 10, 20, and 30 sites. 
At each location, the true $\rho_{\text{t}}^s$ is perturbed by adding $\text{Unif}(-0.05, 0.05)$ noise, making the clustering task more challenging.
We now have a separate t-copula correlation parameter for each of the three clusters, denoted by $\rho_{\text{t}_1}, \rho_{\text{t}_2}$, and $\rho_{\text{t}_3}$.
To reduce the dimensionality of the grid of considered copula parameter values, a constant $\rhou{Gauss} = 0.5$ is used across all sites and experiments. 

\begin{figure}[htb]
    \centering
    \includegraphics[width = 0.7\linewidth]{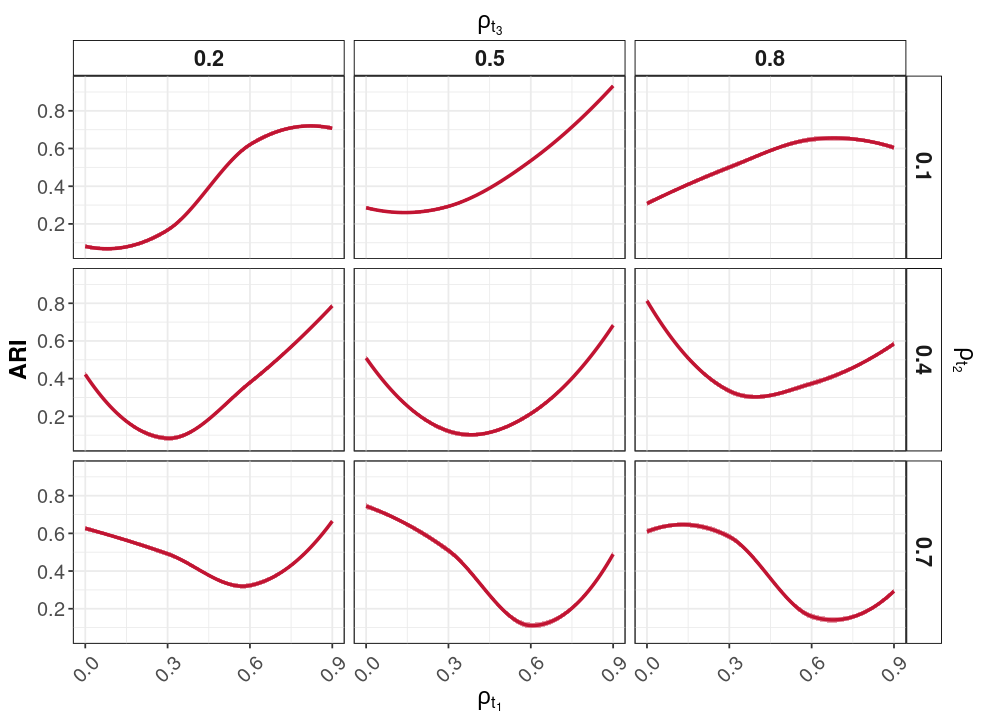}
    \caption{
      Evaluation of clustering performance for $d = 2$ variables and three clusters of 60 locations for simulations from a mixture of Gaussian and t-copulas.
        The true x-axis and facet labels show the t-copula correlation parameters,  $\rho_{\text{t}_1}, \rho_{\text{t}_2}$, and $\rho_{\text{t}_3}$, for each of the three true clusters, for a Gaussian copula correlation of $\rhou{Gauss} = 0.5$. 
    The smoothed lines show the average ARI across bootstrap samples, with smoothing performed using a generalised additive model. 
    }
   \label{fig:03_realistic}
\end{figure}

The results of 500 repetitions of this simulation study are shown in Figure~\ref{fig:03_realistic}.
Similarly to the previous simulation studies, our clustering method performs well, particularly when the differences in t-copula correlation parameters between clusters are larger. 
For example, the best performance is observed where $\rho_{\text{t}_1} = 0.9, \rho_{\text{t}_2} = 0.1$, and $\rho_{\text{t}_3} = 0.5$, as the clusters are most distinct, while conversely, performance is worst for $\rho_{\text{t}_1} = 0.6, \rho_{\text{t}_2} = 0.4$, and $\rho_{\text{t}_3} = 0.5$. 

\section{Application to Irish meteorological data}
\label{sec:application}
We now apply our extremal clustering method to meteorological observations from the Republic of Ireland. 
In Section~\ref{subsec:app_data}, we describe the precipitation and wind speed measurements used in our analysis and present exploratory analysis (using the $\chi$ statistic from Equation~\eqref{eq:chi}) on the extremal dependence between the two variables. 
In Section~\ref{subsec:app_ce}, we discuss fits of the conditional extremes (CE) model, and, in Section~\ref{subsec:app_clust}, we apply our extremal clustering approach. 

\subsection{Ireland meteorological data}
\label{subsec:app_data}

\begin{figure}[htb]
    \centering
    \includegraphics[width = 0.8\linewidth]{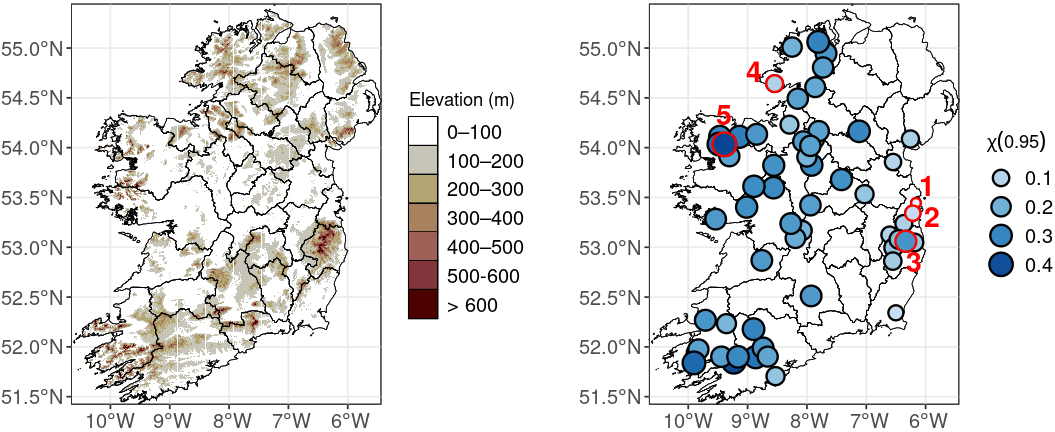}
    \caption{
    Left: elevation (m) profile of Ireland.
    Right: Estimated $\chi(0.95)$ between precipitation and wind speed at all 59 locations. Darker colours and larger points indicate stronger asymptotic dependence.
    Locations in red are (1) Malahide Castle, Dublin, (2) Ringsend, Dublin, (3) Glenmacnass, Wicklow, (4) Kilcar, Donegal, and (5) Derryhillagh, Mayo.
}
   \label{fig:chi}
\end{figure}

We analyse the joint extremal behaviour of precipitation and wind speed at $D=59$ sites across the Republic of Ireland from 1990 to 2020, inclusive. 
Precipitation and wind speed data are obtained from the Met Éireann archive \citep{metHistoricalData}, and ERA5 Reanalysis \citep{Hersbach2020}, respectively. 
Met Éireann data are irregularly spaced over space, while ERA5 provides spatial averages on a regular $0.25^{\circ} \times 0.25^{\circ}$ grid.
We match each spatial site to the grid cell in which it lies. 
The dataset spans 11,323 consecutive daily observations.

Following \citet{Vignotto2021}, we focus on the winter months, October to March, and aggregate precipitation into weekly sums and daily wind speed maxima into weekly averages.
The weekly temporal scale is motivated by the synoptic timescale of extratropical storm systems, which may persist over several days \citep{Bengtsson2009}, and helps reduce short-term temporal dependence not accounted for in our model.
Weeks with no precipitation are excluded, resulting in 5,022 observations across all $D = 59$ sites. 

As the ERA5 wind speed data are available on a relatively coarse $0.25^{\circ} \times 0.25^{\circ}$ grid, some sites fall within the same ERA5 grid cell, and therefore share identical wind speed time series.
To assess the impact of this on the following clustering results, we perform a sensitivity analysis in which only one representative site from each overlapping grid cell was retained.
The resulting clustering solution is very similar to the following, obtained using all sites (Figure~\ref{fig:clust_sol}, left panel), indicating that the substantive conclusions of this analysis are robust to this overlap (see Figure~\ref{fig:overlapping_sites} in the Supplementary Material).

We explore the extremal dependence structure of the data using the $\chi$ statistic introduced in Equation~\eqref{eq:chi}.
Figure~\ref{fig:chi} shows empirical estimates of $\chi(0.95)$ at all 59 sites.
An elevation profile of Ireland is also shown in Figure~\ref{fig:chi}, which may help further explain some of this spatial structure.
Non-zero values are observed at almost all sites, indicating positive extremal dependence at high (but finite) levels between precipitation and wind speed. 
In practical terms, this suggests that extreme wind speeds tend to be accompanied by high precipitation, and vice versa. 
A clear spatial pattern emerges: extremal dependence is strongest in the south west, with the highest values along the Atlantic coast, notably in counties Kerry, Galway, Mayo, and Donegal, and weaker along the east coast, in Dublin, Wexford, and Louth. 
The highest $\chi(0.95)$ value (0.414) is observed at Derryhillagh in Mayo (Figure~\ref{fig:chi}, location (5)), while the lowest (0.0005) is at Malahide Castle in Dublin (Figure~\ref{fig:chi}, location (1); hereafter `Malahide').

Some spatial outliers are also apparent. 
Kilcar in Donegal (Figure~\ref{fig:chi}, location (4)) has a relatively low $\chi(0.95)$ value (0.109), despite its west-coast location, while some sites in county Wicklow show unusually high values for the east, particularly Glenmacnass (Figure~\ref{fig:chi}, location (3); 0.268).
The latter may be related to the influence of the Wicklow Mountains, as reflected in the elevation profile (Figure~\ref{fig:chi}). 
It may be of interest to investigate if these outliers in $\chi(0.95)$ are reflected in the CE model parameter estimates, and whether clustering can identify them.
These results indicate distinct regional patterns of extremal dependence, as well as notable outliers, motivating the use of our extremal clustering framework to characterise and interpret this spatial structure more formally. 

\subsection{Conditional extremes estimates}
\label{subsec:app_ce}

We transform the precipitation and wind speed data, separately at each location, to Laplace margins, with the empirical rank transform replacing $F_{X_i}$ in Equation~\eqref{eq:laplace}.
To select an appropriate quantile level $q$ for fitting the CE model (see Section~\ref{subsec:ce_model}), we estimate the parameters $(\alpha_{\mid i}, \beta_{\mid i})$ at a sequence of levels $q$, to identify the lowest $q$ beyond which parameter estimates appear stable.
Estimates are generally stable for both models, i.e.,\@ precipitation conditional on wind speed and vice versa, down to $q = 0.85$, with similar maximum likelihood estimates across 500 bootstrap samples for $(\alpha_{\mid i}, \beta_{\mid i})$.
Threshold stability plots for the CE fits at the five locations highlighted in Figure~\ref{fig:chi} are included in Figures~\ref{fig:boot_quantiles_malahide} - \ref{fig:boot_quantiles_derryhillagh} of the Supplementary Material.
Therefore, we fix $q = 0.85$ for the subsequent analysis.
In Figure~\ref{fig:clust_sens}, we show that our estimated extremal clustering is robust to the choice of $q$, with estimates shown for $q = 0.85, 0.88$, and $0.90$.

To further assess the adequacy of the CE model fits, we examined diagnostic plots of the fitted residuals at the five locations highlighted in Figure~\ref{fig:chi}, including normal QQ plots of the standardised residuals for both conditioning directions (see Figure~\ref{fig:diag_app} in the Supplementary Material).
Overall, the diagnostics suggest that the working Gaussian assumption is reasonable for the majority of locations considered, although some departures from normality are observed in the tails, as expected for finite threshold levels.

\begin{figure}[htb]
    \centering
    \includegraphics[width = 0.8\linewidth]{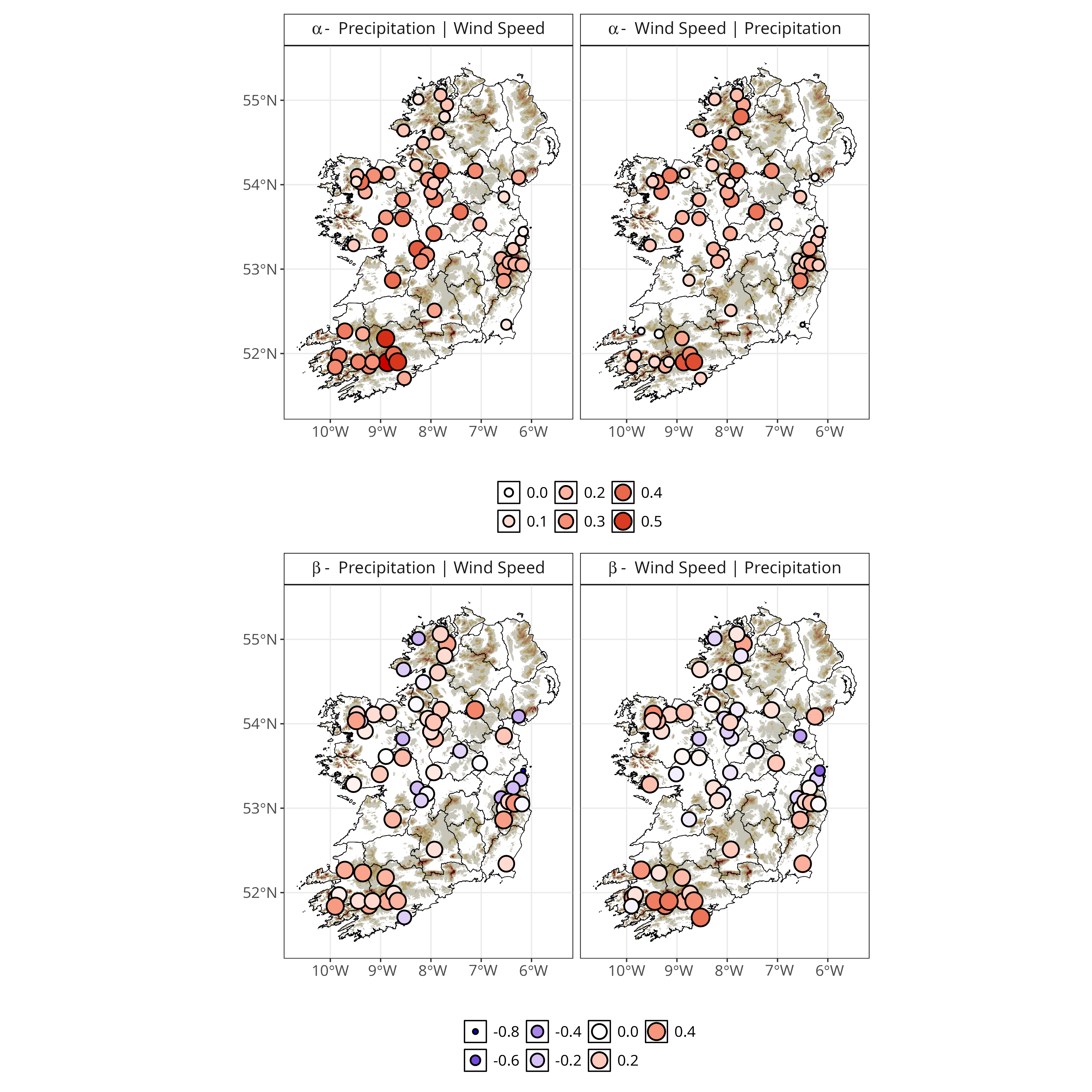}
    \caption{
        Maps of site-wise estimated model parameters $\alpha_{\mid i, s}$ (top) and $\beta_{\mid i, s}$ (bottom) for precipitation conditional on wind speed (left) and wind speed conditional on precipitation (right).
        The colour scale is the same for both variables, with darker red and larger points indicating larger positive values of $\alpha$ and $\beta$, and darker blue and smaller points indicating smaller negative values.
   }
   \label{fig:a_b_vals}
\end{figure}

The estimated site-wise CE parameters are shown in Figure~\ref{fig:a_b_vals}.
Although interpretation is challenging due to variability in the estimates, as with the $\chi(0.95)$ estimates (Figure~\ref{fig:chi}), we observe some evidence of increasing $\alpha_{\mid i}$ values for both variables towards the west and south.
To mitigate uncertainty and aid interpretation, we cluster the locations using our extremal clustering approach. 

\subsection{Clustering}
\label{subsec:app_clust}

We cluster the locations using the aggregated dissimilarity matrix $M$ as defined in Equation~\eqref{eq:aggregate_matrix}, and separately for each individual matrix $M^{(i)}, i = 1, 2$, as in Equation~\eqref{eq:distance_matrix}.
The number of clusters $k$ is chosen via the method in Section~\ref{subsec:clustering}, using the elbow plot of the $\mathrm{TWD}$. 
For all three clustering scenarios, clear elbows are observed at $k = 3$.
The elbow plots are provided in Figure~\ref{fig:elbow_plots} of the Supplementary Material. 

\begin{figure}[htb]
    \centering
    \includegraphics[width = 0.8\linewidth]{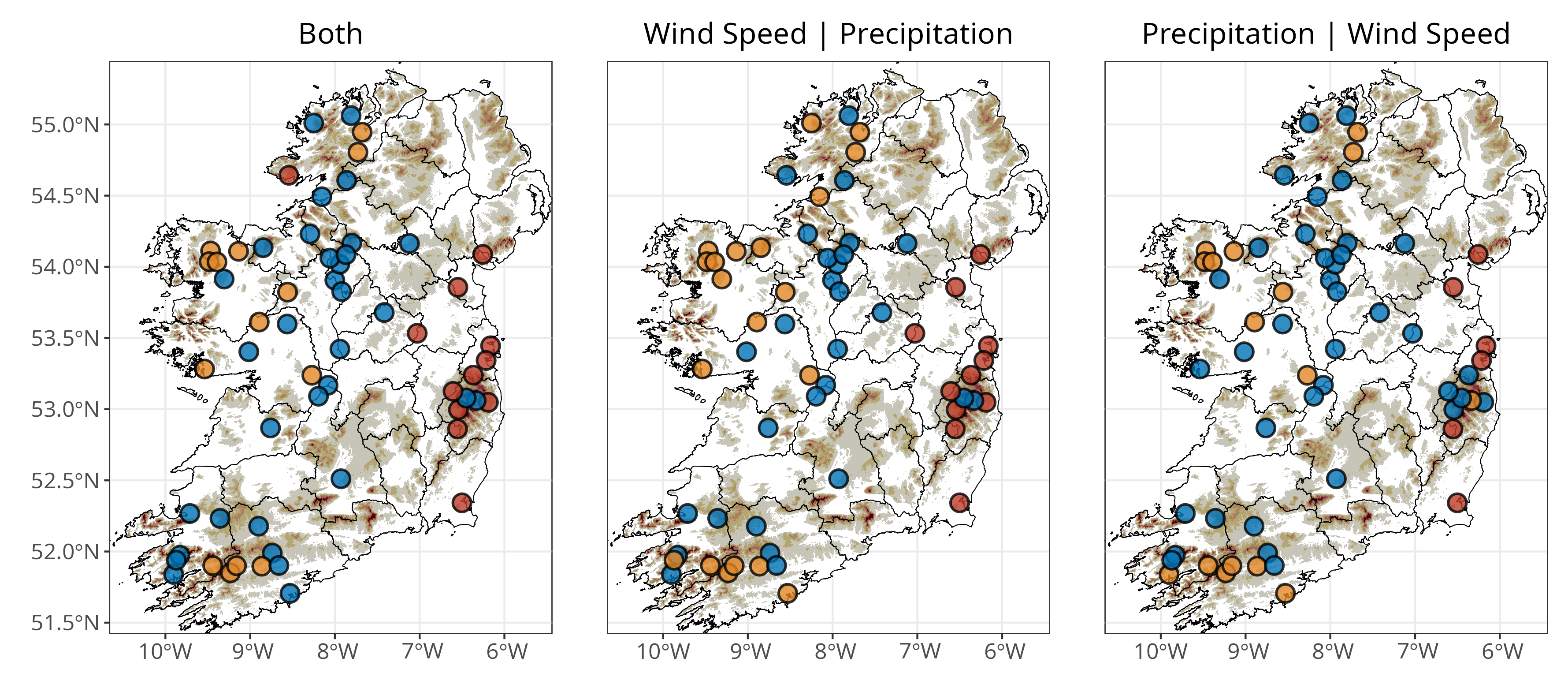}
    \caption{
    Extremal cluster estimates using the dissimilarity matrix for wind speed conditional on precipitation (centre), precipitation conditional on wind speed (right), and the combined dissimilarity matrix (left). 
    Sites are coloured by cluster label.
}
\label{fig:clust_sol}
\end{figure}

The resulting cluster estimates are displayed in Figure~\ref{fig:clust_sol}.  
Across all three choices of dissimilarity matrix, the estimated clusters are broadly consistent, with three spatially distinct groups emerging.  
This spatial coherence is notable given that no explicit spatial information was used in model fitting or in the clustering.
Instead, the clusters are determined entirely by similarities in the fitted CE models through the proposed dissimilarity measure, suggesting that the CE parameters and residual distributions capture meaningful spatial structure in the extremal dependence of the data.

When interpreting the clusters, we observe a clear contrast between the east and west coasts, separated by a central cluster.  
For precipitation conditional on wind speed, the clustering is somewhat less distinct, with the central cluster extending more extensively into parts of Galway, Mayo, and Donegal on the west coast.  
There are also notable outliers in the clusters, which we further discuss below. 

\begin{figure}[htb]
    \centering
    \includegraphics[width = 1.0\linewidth]{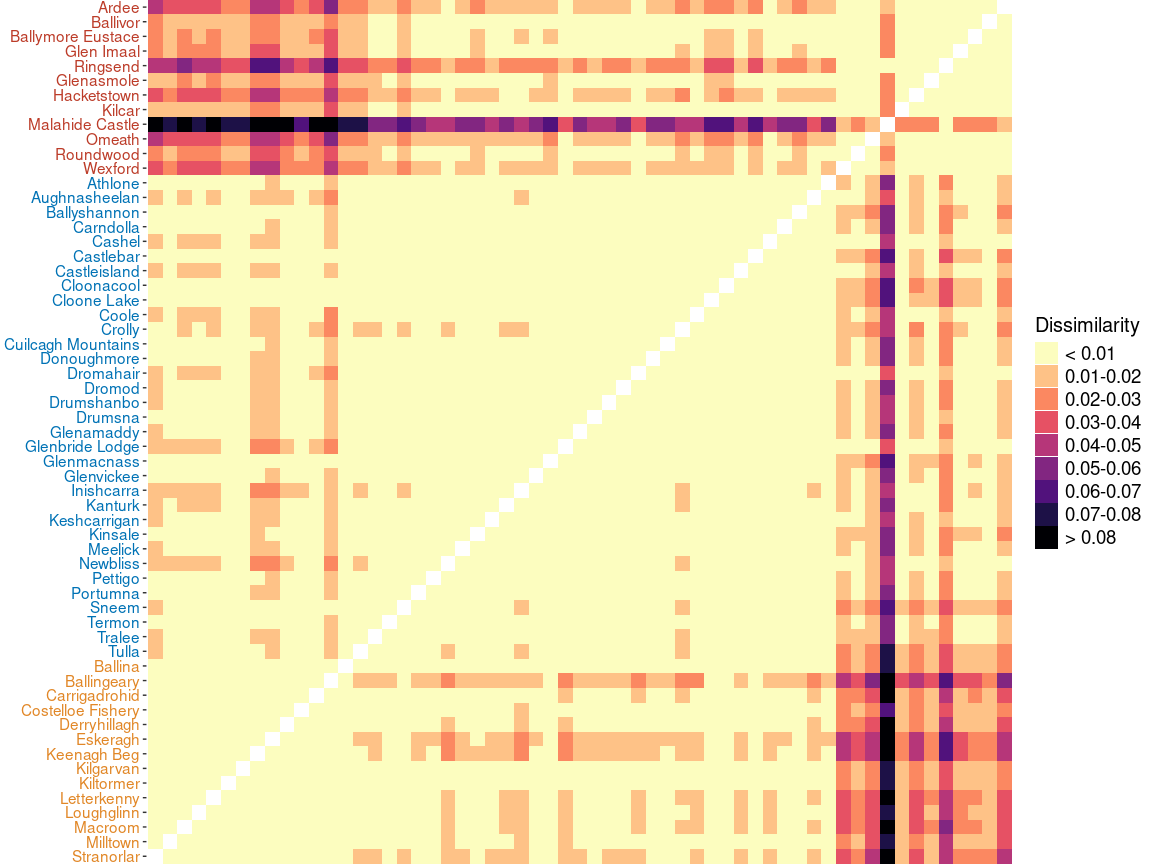}
    \caption{
    Heatmap of the estimated aggregated dissimilarity matrix $M$.
The names of each site are coloured and ordered according to their estimated cluster labels, matching that of Figure \ref{fig:clust_sol}. 
The dissimilarity colour scale is shown on the right, with darker colours indicating larger values.
}
\label{fig:dist_heatmap}
\end{figure}

\begin{figure}[htb]
    \centering
    \includegraphics[width = 0.85\linewidth]{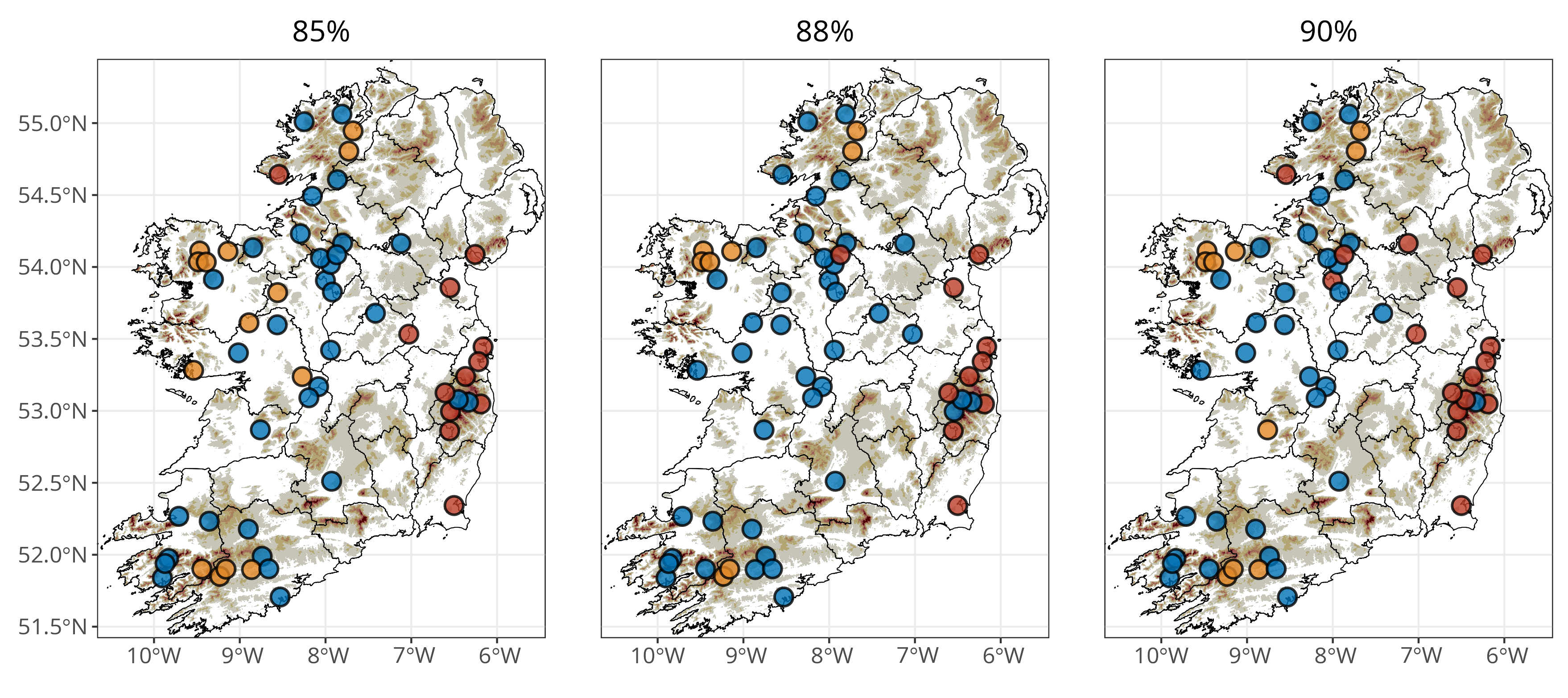}
    \caption{
    Estimated extremal clusters with the conditional extremes model parameters estimated at $q = 0.85, q = 0.88$ and $q = 0.9$ quantiles, using the aggregated dissimilarity matrix. 
    Sites are coloured by cluster label. 
}
\label{fig:clust_sens}
\end{figure}

In Figure~\ref{fig:dist_heatmap}, we visualise the dissimilarity matrix for the aggregated clusters (see left panel of Figure~\ref{fig:clust_sol}).
The clusters are quite distinct, with the dissimilarity between clusters being much larger than within-cluster dissimilarity, which rarely exceeds 0.01.
Within the CE framework, small pairwise dissimilarities correspond to sites having similar estimated CE parameter values and residual distributions, and hence similar conditional extremal dependence behaviour between precipitation and wind speed.
The particularly small within-cluster dissimilarities observed for the central cluster therefore suggest that these sites form a relatively homogeneous group in terms of their extremal dependence structure.
In particular, the difference between the eastern and western clusters is quite pronounced, with the intermediate central cluster acting as somewhat of a `buffer' between the two.
In contrast, the western and central clusters seem to be quite close, in that we observe many small inter-cluster pairwise dissimilarities between them. 
This indicates that allocation between the two is not always clear-cut and for several sites is nearly interchangeable, as observed in  the variable-specific estimated clusters in Figure~\ref{fig:clust_sol} and in the quantile sensitivity analysis in Figure~\ref{fig:clust_sens}, where membership between the two often switches. 
Also, if we only fit two clusters, we find that most of the central cluster members are assigned to the western cluster, reflecting this less obvious separation between the two compared to the more markedly different eastern cluster. 
A noticeable outlier in the dissimilarity matrix is Malahide, which we previously identified as having the lowest $\chi(0.95)$ value  in Figure~\ref{fig:chi}. 
Even within its own cluster, Malahide shows a relatively high dissimilarity to the other sites. 
Indeed, if we set $k=4$, Malahide becomes its own cluster (clustering results for $k=2$ and $k=4$ are provided in Figure~\ref{fig:cluster_dqu_k_2_4} of the Supplementary Material).
Malahide also has the lowest mean weekly precipitation of any site; the second lowest is Ringsend, also in county Dublin (Figure~\ref{fig:chi} location (2)), which is the site closest to Malahide in terms of our dissimilarity measure. 
Finally, Malahide has very low $\chi(0.95)$ (Figure~\ref{fig:chi}) and $\alpha_{\mid i}$ (Figure~\ref{fig:a_b_vals}) estimates. 
Together, this may explain why we observe such outlying behaviour for Malahide.
Finally, Kilcar in County Donegal is the member of the eastern cluster with the smallest dissimilarity to the western cluster members.
This may reflect its geographic location in northwest Ireland, despite it's low $\chi(0.95)$ value (Figure~\ref{fig:chi}) relative to other west-coast sites.

We perform a sensitivity analysis of the estimated clusters (with the aggregated dissimilarity matrix) to the choice of quantile level $q \in \{0.85, 0.88, 0.90\}$, in Figure~\ref{fig:clust_sens}.
The estimated clusters are similar across the three values of $q$, with some small differences. 
Noticeably, the earlier identified outlier at Kilcar in county Donegal is assigned to the eastern cluster when $q = 0.85$ and $q = 0.90$, but to the central cluster when $q = 0.88$.
Also, there are more sites in the western cluster when $q = 0.85$, with some sites moving to the central cluster for $q = 0.88$.
Interestingly, some of these sites then move back to the western cluster for $q = 0.90$.
Apart from these observations, the estimated clusters are generally quite similar across the three quantile levels, and so we conclude that our clustering method is relatively robust to the choice of quantile level.

For comparison, we additionally applied the clustering methodology of \citet{Vignotto2021} to the same dataset (see Appendix~\ref{sec:vignotto_compare} of the Supplementary Material). 
While both methods produced qualitatively similar large-scale spatial clustering patterns, particularly in the separation between eastern, central and western regions of Ireland, the finer site-level cluster assignments differ more substantially.
Furthermore, the \citet{Vignotto2021} approach exhibited greater sensitivity to the choice of extremal threshold level.

\section{Discussion}
\label{sec:discuss}

In this paper, we introduced a novel method for clustering multivariate extremes.
We proposed the expected extremal $\JSGa{}$ divergence as a measure of dissimilarity between multivariate conditional extremes (CE) models \citep{Heffernan2004}.
We leveraged the commonly-applied working Gaussian assumption to derive a closed-form expression for the $\JSGa{}$ divergence.
Efficient k-medoids clustering was then performed on the dissimilarity matrix derived from this measure. 
In our simulation study, a bivariate Gaussian copula example illustrated that our clustering method reduces parameter uncertainty and bias when the CE model is re-fitted with data pooled post-clustering.
Using mixtures of Gaussian and $t$-copulas, we showed that our method outperforms the method of \citet{Vignotto2021} across a range of bivariate extremal dependence scenarios.
Our method was also shown to be uniquely extendable to $d > 2$ dimensions (see Section~\ref{sec:sim_extension} in the Supplementary Material). 
We applied our extremal clustering method to Irish meteorological data, consisting of 59 locations with weekly rainfall and wind speed observations.
Without explicitly incorporating spatial information, three clusters with distinct spatial patterns were identified, with well separated clusters representing the east and west coasts and the centre of Ireland.

One limitation of our Irish meteorological application is that precipitation and wind speed data were obtained from different sources, namely observations at Met Éireann sites and ERA5 reanalysis products, respectively.
This choice was motivated by data availability, as long-term observational wind speed measurements were not available at the sites considered in this study.
We retained the station-based precipitation measurements, rather than using ERA5 precipitation estimates, because gridded reanalysis products may smooth localised rainfall extremes, potentially leading to attenuation of extremal behaviour \citep{Risser2019, Glawion2025}. 
Nevertheless, combining observational and reanalysis datasets may introduce discrepancies due to differences in spatial and temporal resolution, measurement uncertainty, and different data-generation methodologies.
In particular, because the ERA5 wind speed data are available on a relatively coarse $0.25^{\circ} \times 0.25^{\circ}$ grid, some sites shared identical wind speed time series.
However, a sensitivity analysis presented in Section~\ref{sec:overlapping} of the Supplementary Material showed that retaining only one representative site from each overlapping ERA5 grid cell produced very similar clustering results, suggesting that the substantive conclusions of the application are robust to this overlap. 
More broadly, the primary aims of the Irish case study were to demonstrate the proposed extremal clustering methodology in a realistic environmental setting, rather than to provide definitive inference on the physical extremal dependence structure between precipitation and wind speed across Ireland.
A more detailed investigation of the impact of alternative meteorological data products and harmonisation strategies would therefore be a valuable direction for future applied work.

While we focused on a spatial application in this work, our method is more generally applicable to any setting where clustering of multivariate extremal dependence is desired.
For instance, it could be applied to clinical trial or public health data to cluster patients based on the tail dependence structure of their symptoms or biomarker measurements.
A neuroscience application was explored in \citet{Talento2025}, where the CE framework was used to highlight differences in the tail dependence of electroencephalogram (EEG) signals for seizure-prone and seizure-absent subjects. 
However, they used pre-defined clusters. Hence, we may expect our approach to be useful in similar applications, to identify subjects who have undergone extreme brain events, e.g., seizures.
In financial applications, we could cluster time periods according to the extremal dependence of log-returns of different assets, offering potential insights for portfolio management and risk assessment.

While the residual Gaussian assumption on which our method relies may not hold for all datasets, our clustering approach showed good performance for all settings in Section~\ref{sec:sim}.
A possible alternative is to employ the generalised and asymmetric multivariate Gaussian distribution used in \citet{Farrell2025}. 
While this does not permit a closed-form divergence, our extremal clustering method could still be used, with costly numerical techniques used to estimate the dissimilarity matrix.
Indeed, if choosing to forgo this Gaussian assumption, it may be preferable to use the standard Jensen-Shannon divergence in place of the $\JSGa{}$ divergence, since the former is bounded \citep{Thiagarajan2025}.

Another important limitation of our method is in its treatment of uncertainty.
We do not propagate uncertainty in the CE model estimates through to the clustering. 
To account for this, one possible approach is to use the bootstrap method of \citet{Heffernan2004} to generate samples of CE model estimates, and compute our dissimilarity matrix for each sample. 
We could then produce a distribution of dissimilarity matrices, and use these to compute a distribution of estimated clusters.
However, this would be computationally expensive, and so would compromise one of the main advantages of our method, namely its computational efficiency.

In this work, we implicitly assume spatial stationarity in the extremal dependence structure, as we fit a separate CE model at each location.
Extensions of the CE model have been proposed, such as those incorporating covariates \citep{Richards2023} and spatio- and spatio-temporal representations of the CE parameters \citep{Simpson2021, Wadsworth2022}.
These could be incorporated into our clustering framework, but care must be taken to ensure that the residual Gaussian assumption still holds, or that an appropriate alternative dissimilarity is used, as discussed above.
Another approach to dealing with the uncertainty in the clustering solution would be to adopt a Bayesian perspective.
By placing priors on the CE model parameters, we could obtain posterior distributions for these parameters, which could be used to compute a distribution of dissimilarity matrices and clustering solutions.
Approaches such as that of \citet{Rohrbeck2021} have used a Reversible-Jump MCMC scheme to perform Bayesian clustering for spatial extremes, which does not require the number of clusters to be specified in advance.

\backmatter

\bmhead{Supplementary information}

%
%

Supplementary information accompanies this article and contains additional theoretical results, simulation diagnostics, sensitivity analyses, and supporting results for the proposed methodology and application.

\bmhead{Acknowledgements}

%

Patrick O'Toole is supported by a scholarship from the EPSRC Centre for Doctoral Training in Statistical Applied Mathematics at Bath (SAMBa), under the project EP/S022945/1.

We acknowledge the use of the ERA5 reanalysis dataset, available from the Copernicus Climate Change Service (C3S) Climate Data Store (CDS) (\url{https://cds.climate.copernicus.eu/}). 
We also acknowledge the use of the Irish meteorological dataset obtained from Met Éireann (\url{https://www.met.ie/climate/available-data/historical-data}).

\section*{Declarations}

%
%

\subsection*{Funding}

Patrick O'Toole is supported by a scholarship from the EPSRC Centre for Doctoral Training in Statistical Applied Mathematics at Bath (SAMBa), under grant EP/S022945/1.

%
%
%
%
%
%
%
%
\subsection*{Data availability}

We acknowledge the use of the ERA5 reanalysis dataset, available from the Copernicus Climate Change Service (C3S) Climate Data Store (CDS) (\url{https://cds.climate.copernicus.eu/}). 
We also acknowledge the use of the Irish meteorological dataset obtained from Met Éireann (\url{https://www.met.ie/climate/available-data/historical-data}).
The processed data used in this study are publicly available on Zenodo at
\url{https://zenodo.org/records/17423842}.

%

\subsection*{Code availability}

The proposed methodology is implemented in the R package \texttt{CeCl}. 
The analysis code used to produce the results in this paper is available from the corresponding author upon reasonable request.

\bibliography{library}

\clearpage

\subfile{supplementary_material}

\end{document}

%% file: supplementary_material.tex
\clearpage

\begin{center}
  {\Large\bfseries
  Supplementary Material for\\[0.4em]
  ``Clustering of Multivariate Tail Dependence Using Conditional Methods''
  }

  %

  \vspace{1.5em}

  {\bfseries Abstract}

  \vspace{0.5em}

  \begin{minipage}{0.85\textwidth}
  This supplementary material contains additional theoretical results,
  diagnostics, sensitivity studies, and comparison plots supporting the
  methodology and application presented in the main manuscript.
  These include a proof of boundedness for the proposed expected divergence
  measure, simulation study diagnostics, threshold stability plots for the
  conditional extremes (CE) model fits, and diagnostic plots assessing the
  adequacy of the CE modelling assumptions.
  Additional sensitivity analyses are provided for the Irish meteorological
  application, including robustness to overlapping ERA5 grid cells and
  alternative choices of the number of clusters.
  We also include further comparisons between our proposed extremal clustering
  methodology and that of \citet{Vignotto2021}.
  \end{minipage}
\end{center}

\vspace{1.5em}

\setcounter{section}{0}
\setcounter{figure}{0}
\setcounter{table}{0}
\setcounter{equation}{0}
\setcounter{theorem}{0}

\renewcommand{\thesection}{S\arabic{section}}
\renewcommand{\thefigure}{S\arabic{figure}}
\renewcommand{\thetable}{S\arabic{table}}
\renewcommand{\theequation}{S.\arabic{equation}}

\section{Proof of Proposition~\ref{prop:bounded}}
\label{sec:proof_boundedness}

We restate the proposition from the main text for completeness.
\begin{proposition}
Assume that $\Sigma_{\mid i,s}$ and $\Sigma_{\mid i,s^*}$ in Equation~\eqref{eq:ce_norm} are positive definite.
Then, for any pair of sites $(s,s^*)$, the expected extremal skew-geometric Jensen-Shannon divergence is bounded, i.e.,
\[
0\le \eJSGa{s,s^*}<\infty.
\]
\end{proposition}

\begin{proof}
To show that the integral is bounded above, we consider the asymptotic behaviour of the integrand.
For convenience, define
\[
A_{i,s}(y)
=
\operatorname{diag}
\left(
y^{\beta_{j\mid i,s}}
:
j\in\{1,\ldots,d\}\smallsetminus\{i\}
\right).
\]
Now write
\[
\boldsymbol{m}(y)
=
\boldsymbol{\alpha}_{\mid i,s}y
+
y^{\boldsymbol{\beta}_{\mid i,s}}
\boldsymbol{\mu}_{\mid i,s},
\qquad
\Omega(y)
=
A_{i,s}(y)
\Sigma_{\mid i,s}
A_{i,s}(y).
\]
and define $A_{i,s^*}(y)$, $\boldsymbol{m}^*(y)$ and $\Omega^*(y)$ analogously for site $s^*$.

Since $\alpha_{j\mid i,s}\in[-1,1]$ and $\beta_{j\mid i,s}\le 1$, each component of $\boldsymbol{m}(y)$ and $\boldsymbol{m}^*(y)$ is $O(y)$ as $y\to\infty$. 
Moreover, because $\Sigma_{\mid i,s}$ and $\Sigma_{\mid i,s^*}$ are positive definite, and because $A_{i,s}(y)$ and $A_{i,s^*}(y)$ are invertible for $y>0$,
\[
\Omega(y)^{-1}
=
A_{i,s}(y)^{-1}
\Sigma_{\mid i,s}^{-1}
A_{i,s}(y)^{-1},
\]
and similarly for $\Omega^*(y)^{-1}$.

Define
$
\beta_{\min}
=
\min\limits_{j\in\{1, \ldots,d\}\smallsetminus\{i\}}
\beta_{j\mid i,s} 
$
and
$
\beta_{\min}^*
=
\min\limits_{j\in\{1, \ldots,d\}\smallsetminus\{i\}}
\beta_{j\mid i,s^*}.
$
Then every entry of $\Omega(y)^{-1}$ has asymptotic behaviour at most of order
\[
O\!\left(y^{-2\beta_{\min}}\right),
\]
and, similarly, every entry of $\Omega^*(y)^{-1}$ is at most of order
$
O\!\left(y^{-2\beta_{\min}^*}\right)
$.

Substituting these orders into the closed-form expression in Equation~\eqref{eq:js_norm_mvn1} yields
\[
\cJSGa{s,s^*}(y)
=
O\!\left(
y^{2-2\min(\beta_{\min},\beta_{\min}^*)}
\right)
+
O(\log y),
\qquad y\to\infty.
\]

Under the standardised Laplace margins used throughout,
\[
h(y\mid Y_{i,s}>u_i)=Ke^{-y},
\qquad y>u_i,
\]
for some constant $K>0$. Hence, the integrand has asymptotic behaviour
\[
\cJSGa{s,s^*}(y)\,h(y\mid Y_{i,s}>u_i)
=
O\!\left(
y^{2-2\min(\beta_{\min},\beta_{\min}^*)}e^{-y}
\right)
+
O\!\left((\log y)e^{-y}\right).
\]

The leading term is analogous to a Gamma density kernel of the form $y^r e^{-y}$, where
\[
r=2-2\min(\beta_{\min},\beta_{\min}^*)\ge 0.
\]
Since exponential decay dominates polynomial growth, we have that
\[
\int_{u_i}^{\infty} y^r e^{-y}\,\mathrm{d}y<\infty
\]
for all $r>-1$. 
Moreover,
\[
\int_{u_i}^{\infty} (\log y)e^{-y}\,\mathrm{d}y<\infty.
\]
Therefore,
\[
\eJSGa{s,s^*}
=
\int_{u_i}^{\infty}
\cJSGa{s,s^*}(y)
h(y\mid Y_{i,s}>u_i)\,\mathrm{d}y
<
\infty, 
\]
as both $\cJSGa{s,s^*}(y)$ and $h(y\mid Y_{i,s}>u_i)$ are non-negative for all $y>u_i$, it follows that $\eJSGa{s,s^*}\ge 0$.
\end{proof}

\section{Simulation study diagnostics}
\label{sec:sim_diag}

\begin{figure}[ht]
    \centering
    \includegraphics[width = 0.9\linewidth]{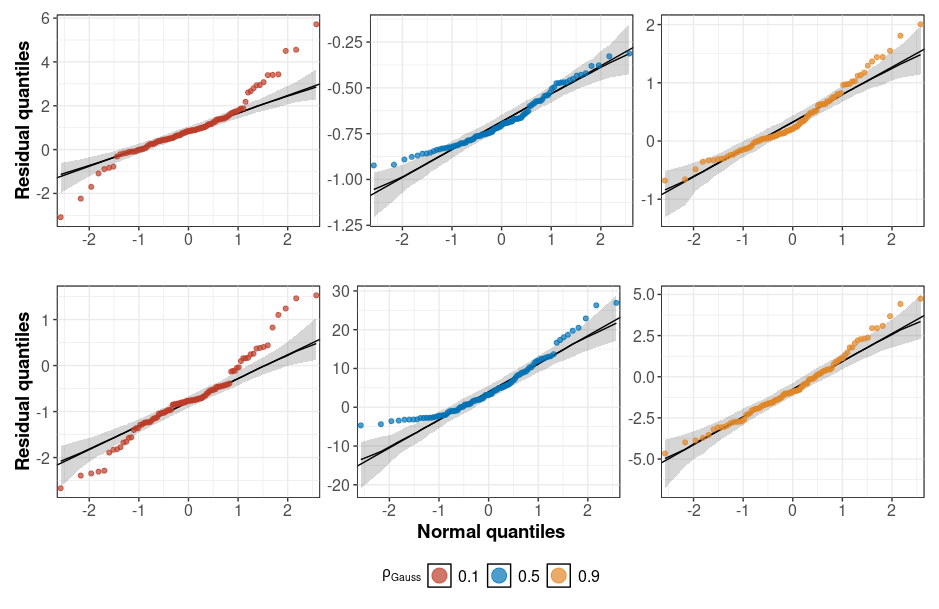}
    \caption{
          Normal quantile-quantile (QQ) plots of the estimated conditional extremes residuals for both conditional models ($X_{1, s} \mid X_{2, s}$ (top) and $X_{2, s} \mid X_{1, s}$ (bottom)) from a single experiment. 
    The solid black line denotes the quartile-based QQ reference line, while the shaded region represents pointwise 95\% simulation envelopes obtained from Gaussian samples.
    }
    \label{fig:diag_gauss}
\end{figure}

Figure~\ref{fig:diag_gauss} shows diagnostic normal quantile-quantile (QQ) plots for the standardised CE residuals $Z_{j\mid i, s}$ for a single experiment from the Gaussian copula simulation setting, described in Section~\ref{subsec:sim_gauss} of the main text. 
Results are shown for both conditional models ($X_{1, s} \mid X_{2, s}$ and $X_{2, s} \mid X_{1, s}$), at three representative sites (one per cluster), using the 0.99 standard Laplace quantile as the conditioning threshold, $u_i$.

\clearpage

\section{Three-dimensional simulation study}
\label{sec:sim_extension}

While the method of \citet{Vignotto2021} is restricted to $d = 2$ dimensions, our approach can be extended to $d \geq 3$ dimensions. 
To illustrate this, we extend the simulation study of Section~\ref{subsubsec:sim_competing_methods} in the main text by considering a three-dimensional mixture copula model.
As in Section~\ref{subsec:sim_mixture} of the main text, observations are generated from a mixture of Gaussian and Student $t$-copulas with Laplace margins, but now with $d = 3$ variables observed at $D = 12$ locations, each with $n = 1000$ observations.
The Gaussian copula component is defined exactly as in the bivariate setting, with a common correlation parameter $\rhou{Gauss}$ shared across all variable pairs and both clusters.

In contrast to the bivariate setting, we consider a collection of trivariate $t$-copula correlation structures $R_{t_1}$ and $R_{t_2}$ for the two clusters. 
Each $R_t$ is a $3 \times 3$ correlation matrix of the form
\[
R_t =
\begin{pmatrix}
1 & \rho_{12} & \rho_{13} \\
\rho_{12} & 1 & \rho_{23} \\
\rho_{13} & \rho_{23} & 1
\end{pmatrix},
\]
where $(\rho_{12}, \rho_{13}, \rho_{23})$ denote the pairwise correlations.
For convenience, we represent each $R_t$ by the vector of its off-diagonal elements.

\begin{figure}[htb]
    \centering
    \includegraphics[width=0.9\linewidth]{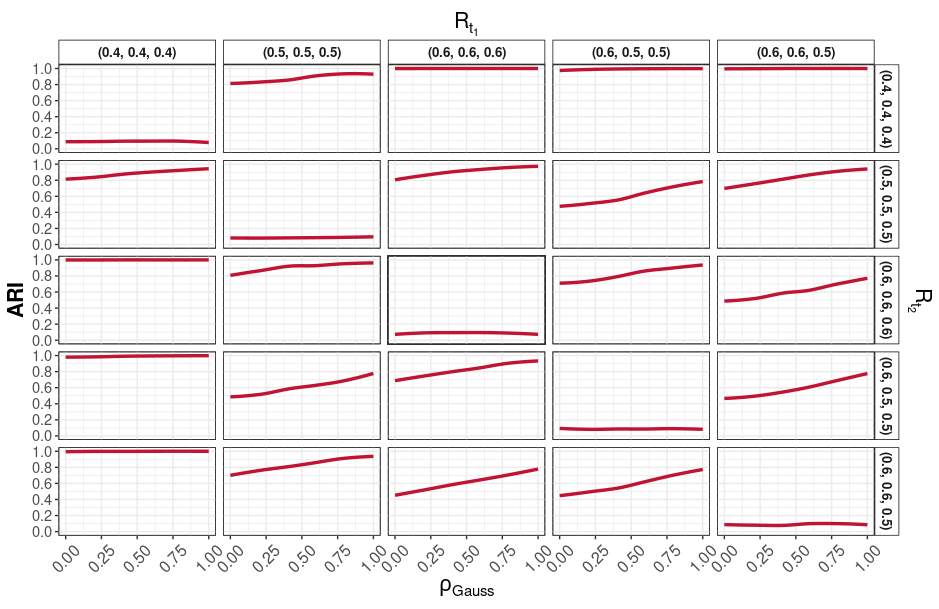}
    \caption{
        Comparison of clustering performance for $d=3$ dimensions and two equally-sized clusters. 
        The x-axis represents the Gaussian correlation parameter $\rhou{Gauss}$ for both clusters, and the facet labels show the $t$-copula correlation structures $R_{t_1}$ and $R_{t_2}$, represented by $(\rho_{12}, \rho_{13}, \rho_{23})$, for the two clusters. 
        The smoothed lines show the average ARI across bootstrap samples, with smoothing performed using a generalised additive model.
    }
   \label{fig:02_3d}
\end{figure}

We construct a set of such correlation structures using values in $\{0.4, 0.5, 0.6\}$, including both homogeneous and heterogeneous configurations (see Table~\ref{tab:rt_patterns}), thereby allowing the extremal dependence structure to vary across variable pairs.
These values are deliberately chosen to be relatively close, so that the resulting cluster-specific dependence structures are similar and the clustering task is non-trivial. 
The Gaussian copula correlation parameter $\rhou{Gauss}$ is varied over the same grid of values
$\rhou{Gauss} \in \{0.1, 0.2, \ldots, 1.0\}$ as considered in Section~\ref{subsubsec:sim_competing_methods} of the main text, and is shared across all variable pairs and both clusters.
Our mixture copula now has three dimensions, and we fit six CE models at each site. 
Clustering is performed using the aggregated divergence matrix $M$ from Equation~\eqref{eq:aggregate_matrix}.
The results in Figure~\ref{fig:02_3d} show that our proposed extremal clustering method performs well in this multivariate setting, with clustering accuracy depending on the similarity between the cluster-specific correlation matrices $R_{t_1}$ and $R_{t_2}$. 
As expected, performance is highest when the extremal dependence structures differ substantially, and more moderate when the differences are subtle, demonstrating that the method is able to exploit multivariate extremal dependence information beyond the bivariate case.
Notably, the Adjusted Rand Index (ARI) along the diagonal, where $R_{t_1} = R_{t_2}$, is close to zero, as the two clusters share the same extremal dependence structure and are therefore indistinguishable, yielding similar performance to random assignment.

\begin{table}[h]
\centering
\begin{tabular}{ll}
\hline
$R_t$ & Description \\
\hline
\multicolumn{2}{l}{\textit{Homogeneous}} \\
$(0.4, 0.4, 0.4)$ & $\rho_{12} = \rho_{13} = \rho_{23}$ (low) \\
$(0.5, 0.5, 0.5)$ & $\rho_{12} = \rho_{13} = \rho_{23}$ (moderate) \\
$(0.6, 0.6, 0.6)$ & $\rho_{12} = \rho_{13} = \rho_{23}$ (high) \\[0.3em]
\multicolumn{2}{l}{\textit{One-high}} \\
$(0.6, 0.5, 0.5)$ & $\rho_{12} > \rho_{13} = \rho_{23}$ \\
\multicolumn{2}{l}{\textit{Two-high}} \\
$(0.6, 0.6, 0.5)$ & $\rho_{12} = \rho_{13} > \rho_{23}$ \\
\hline
\end{tabular}
\caption{Trivariate $t$-copula correlation structures $R_t$, represented by the off-diagonal elements $(\rho_{12}, \rho_{13}, \rho_{23})$ of the correlation matrix.}
\label{tab:rt_patterns}
\end{table}

\section{Threshold stability plots}

Here, we present threshold stability plots for the conditional extremes model fits at the five locations highlighted in Figure~\ref{fig:chi} of the main text, or specifically, Malahide Castle and Ringsend, both in county Dublin, Glenmacnass in county Wicklow, Kilcar in county Donegal, and Derryhillagh in county Mayo.

\begin{figure}[htb]
    \centering
    \includegraphics[width = 0.62\linewidth]{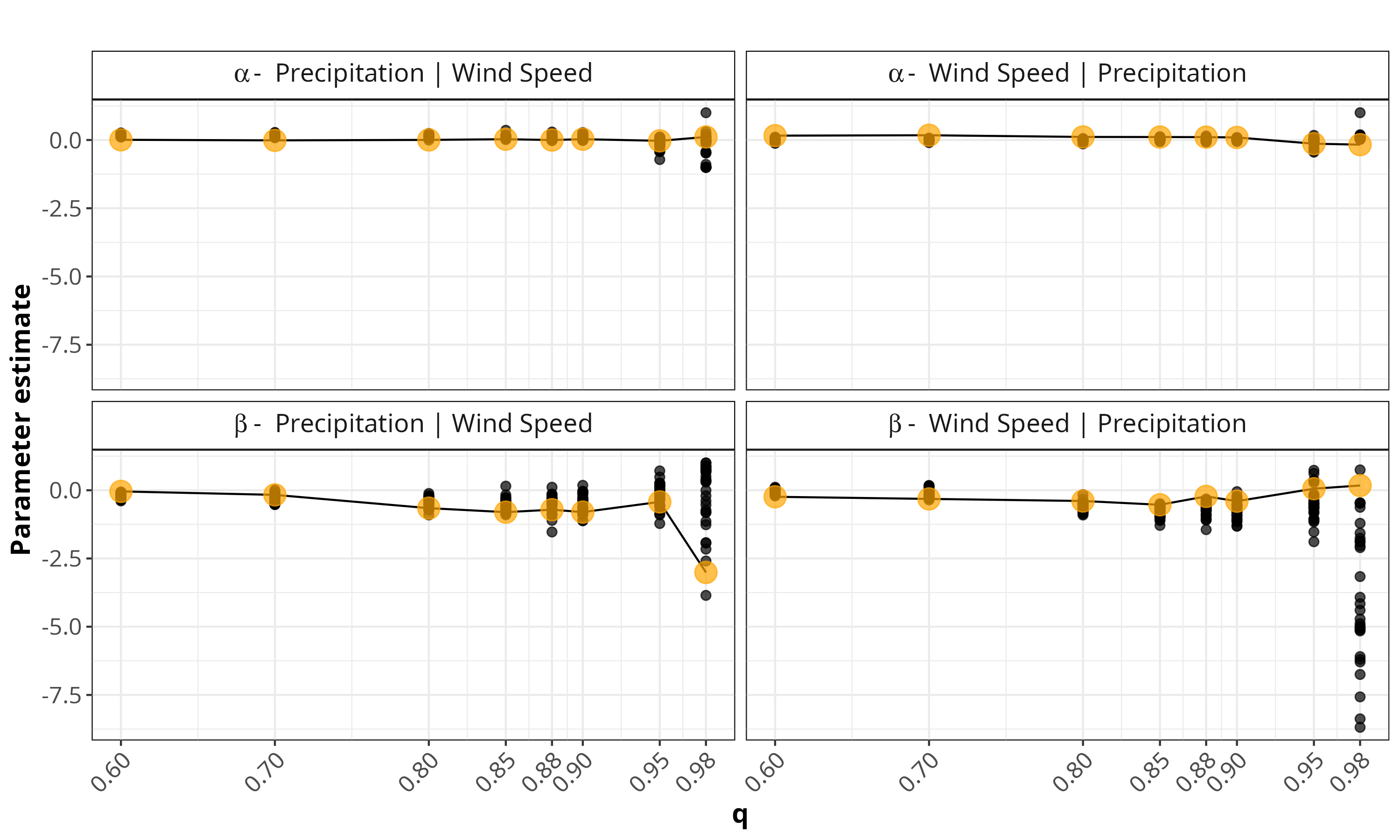}
    \caption{
        Threshold stability plots for estimated model parameters $\alpha_{\mid i, s}$ (top) and $\beta_{\mid i, s}$ (bottom) for precipitation conditional on wind speed (left) and wind speed conditional on precipitation (right) at Malahide Castle, county Dublin.
        Estimates for 500 bootstrap samples at each quantile level $q$ of the standard Laplace distribution are shown as black points, with yellow points indicating the estimates obtained using the full data set at each threshold.
    }
    \label{fig:boot_quantiles_malahide}
\end{figure}

\begin{figure}[htb]
    \centering
    \includegraphics[width = 0.62\linewidth]{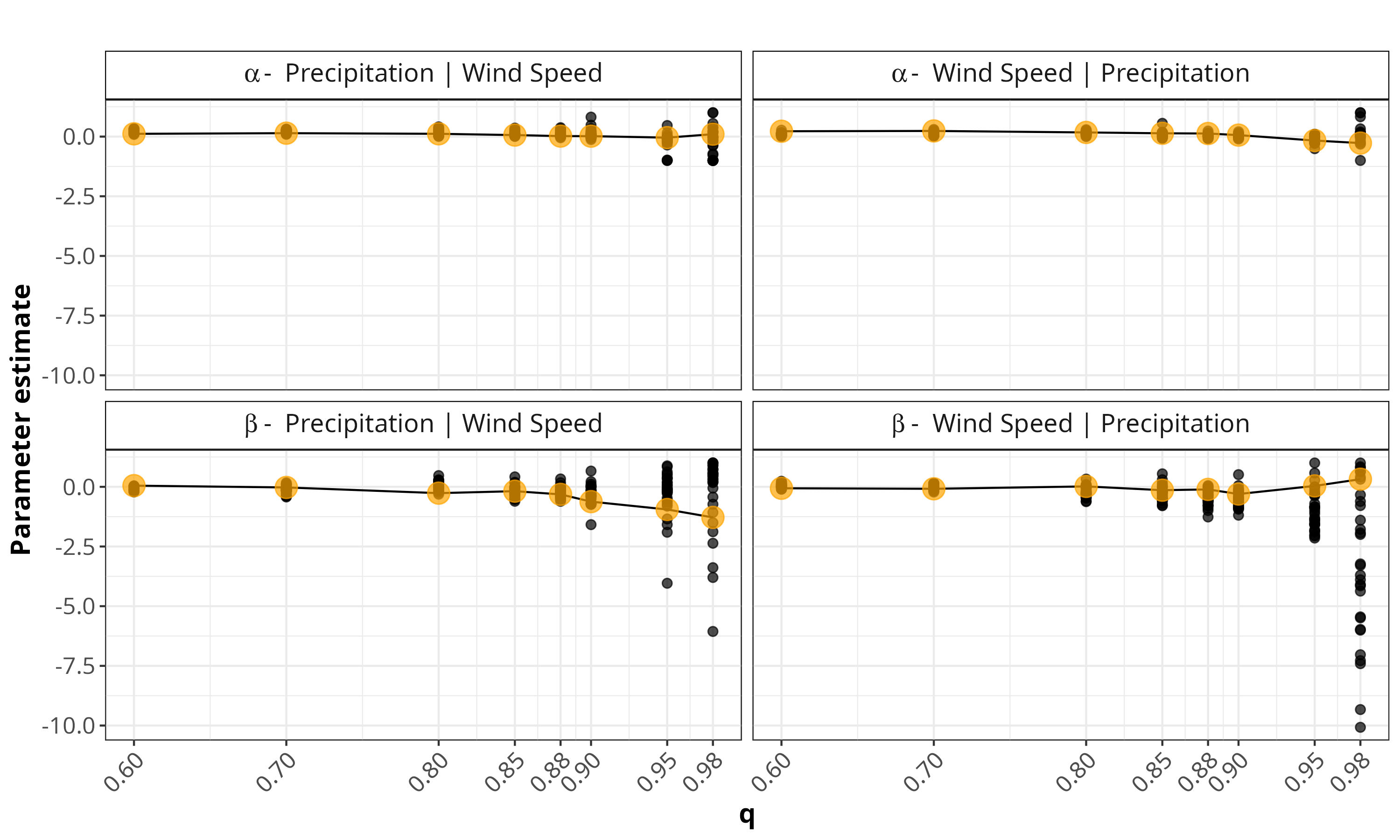}
    \caption{
        Threshold stability plots for estimated model parameters $\alpha_{\mid i, s}$ (top) and $\beta_{\mid i, s}$ (bottom) for precipitation conditional on wind speed (left) and wind speed conditional on precipitation (right) at Ringsend, county Dublin.
        Estimates for 500 bootstrap samples at each quantile level $q$ of the standard Laplace distribution are shown as black points, with yellow points indicating the estimates obtained using the full data set at each threshold.
    }
    \label{fig:boot_quantiles_ringsend}
\end{figure}

\begin{figure}[htb]
    \centering
    \includegraphics[width = 0.62\linewidth]{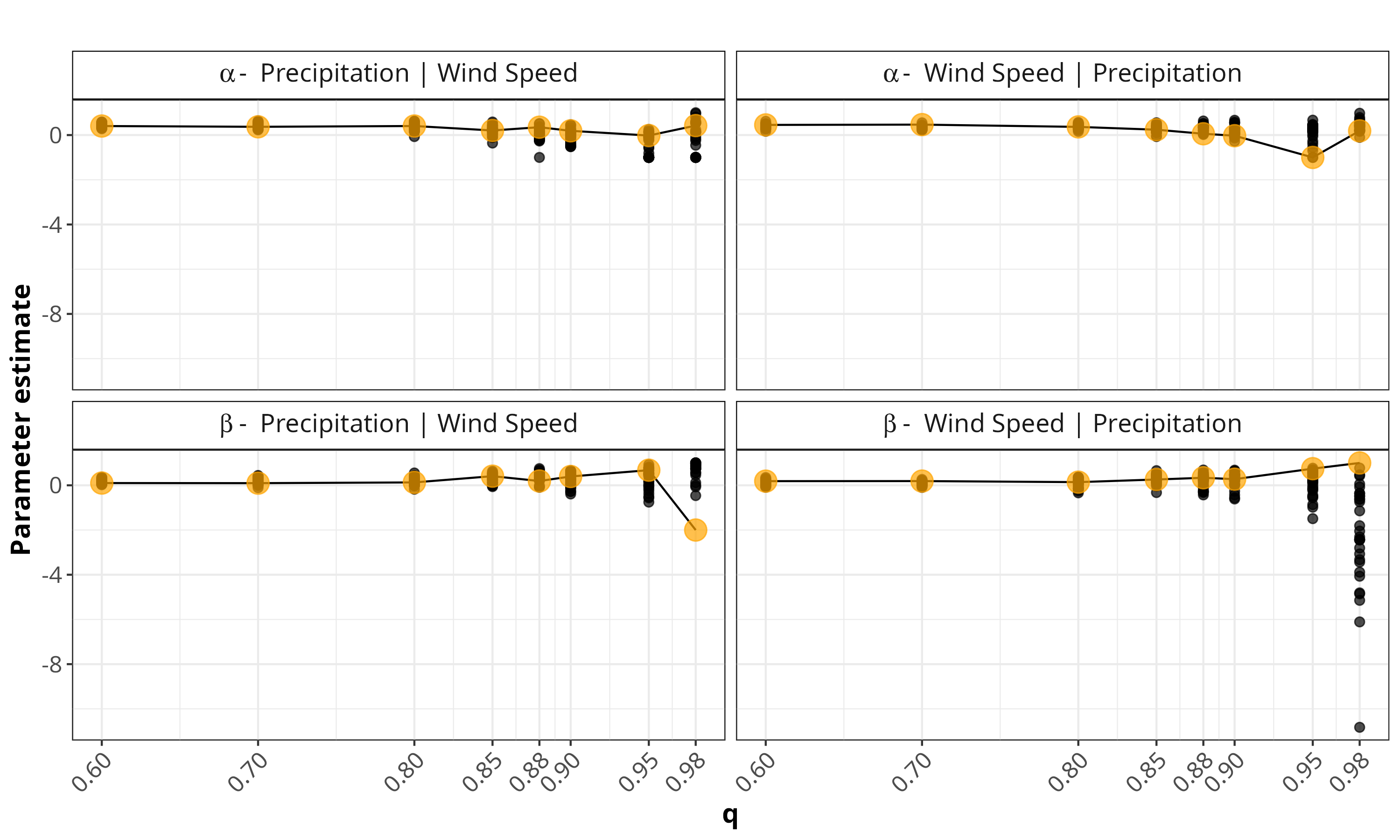}
    \caption{
        Threshold stability plots for estimated model parameters $\alpha_{\mid i, s}$ (top) and $\beta_{\mid i, s}$ (bottom) for precipitation conditional on wind speed (left) and wind speed conditional on precipitation (right) at Glenmacnass, county Wicklow.
        Estimates for 500 bootstrap samples at each quantile level $q$ of the standard Laplace distribution are shown as black points, with yellow points indicating the estimates obtained using the full data set at each threshold.
    }
    \label{fig:boot_quantiles_glenmacnass}
\end{figure}

\begin{figure}[htb]
    \centering
    \includegraphics[width = 0.62\linewidth]{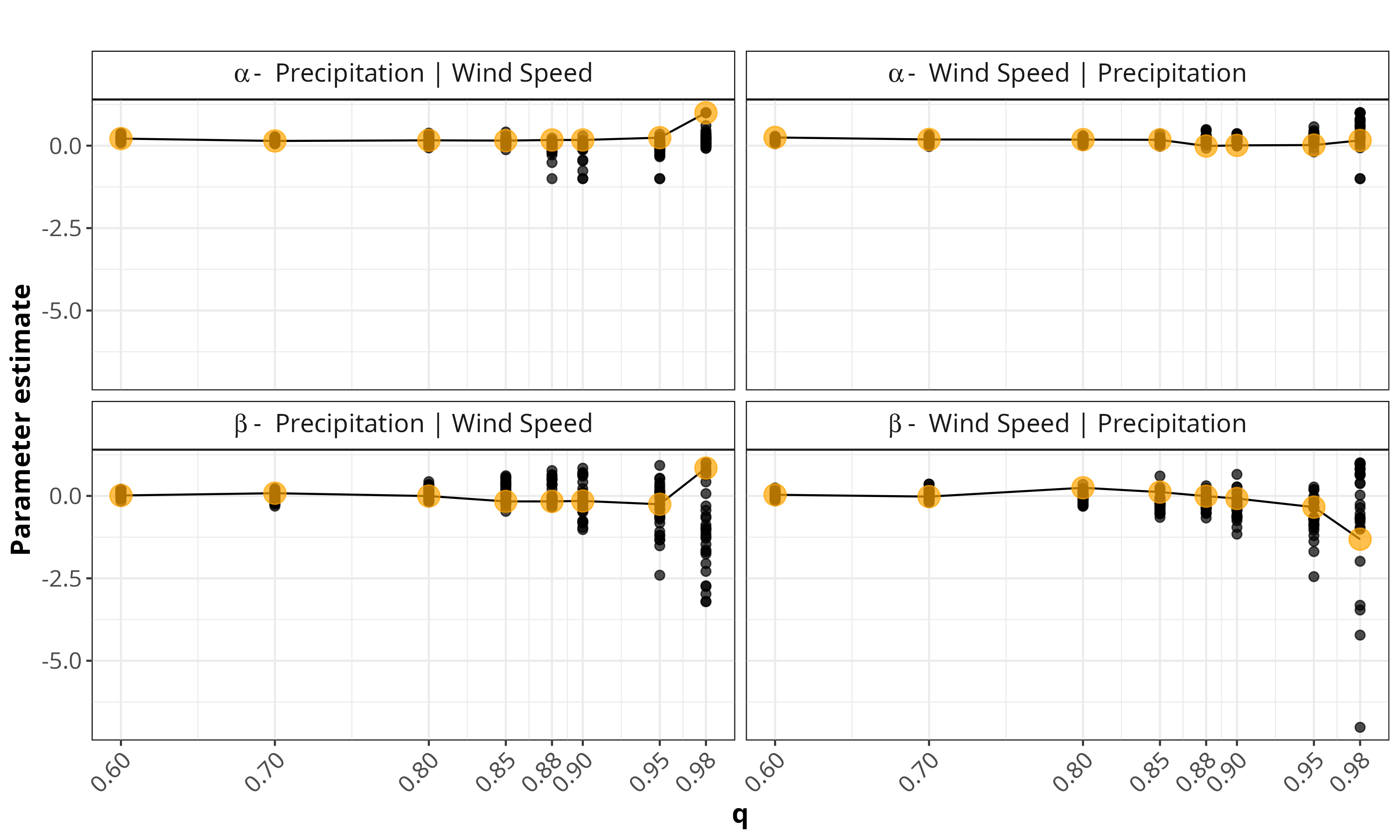}
    \caption{
        Threshold stability plots for estimated model parameters $\alpha_{\mid i, s}$ (top) and $\beta_{\mid i, s}$ (bottom) for precipitation conditional on wind speed (left) and wind speed conditional on precipitation (right) at Kilcar, county Donegal.
        Estimates for 500 bootstrap samples at each quantile level $q$ of the standard Laplace distribution are shown as black points, with yellow points indicating the estimates obtained using the full data set at each threshold.
    }
    \label{fig:boot_quantiles_kilcar}
\end{figure}

\begin{figure}[htb]
    \centering
    \includegraphics[width = 0.62\linewidth]{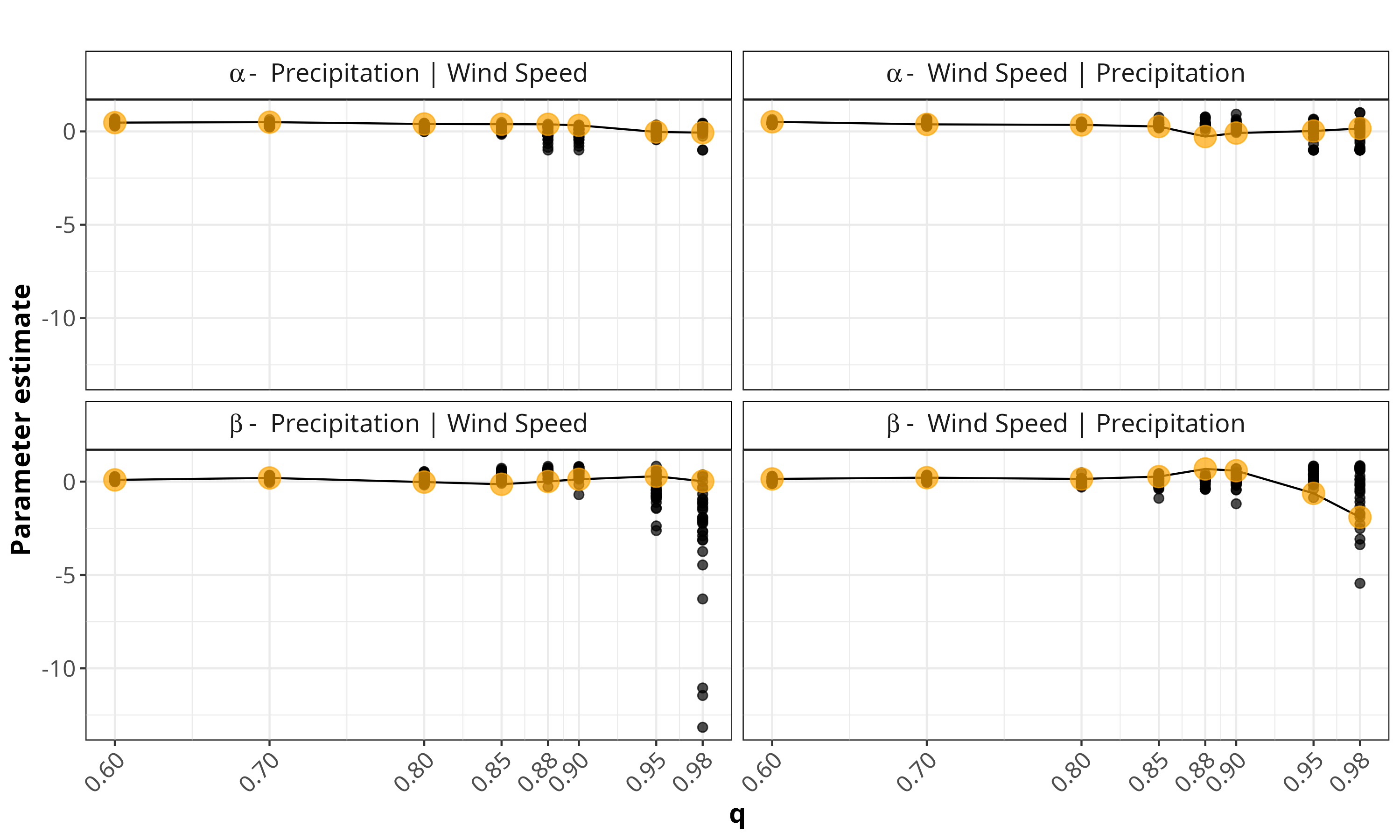}
    \caption{
        Threshold stability plots for estimated model parameters $\alpha_{\mid i, s}$ (top) and $\beta_{\mid i, s}$ (bottom) for precipitation conditional on wind speed (left) and wind speed conditional on precipitation (right) at Derryhillagh, county Mayo.
        Estimates for 500 bootstrap samples at each quantile level $q$ of the standard Laplace distribution are shown as black points, with yellow points indicating the estimates obtained using the full data set at each threshold.
    }
    \label{fig:boot_quantiles_derryhillagh}
\end{figure}

\section{Application diagnostics}
\label{sec:ce_diag}

\begin{figure}[ht]
    \centering
    \includegraphics[width = 0.9\linewidth]{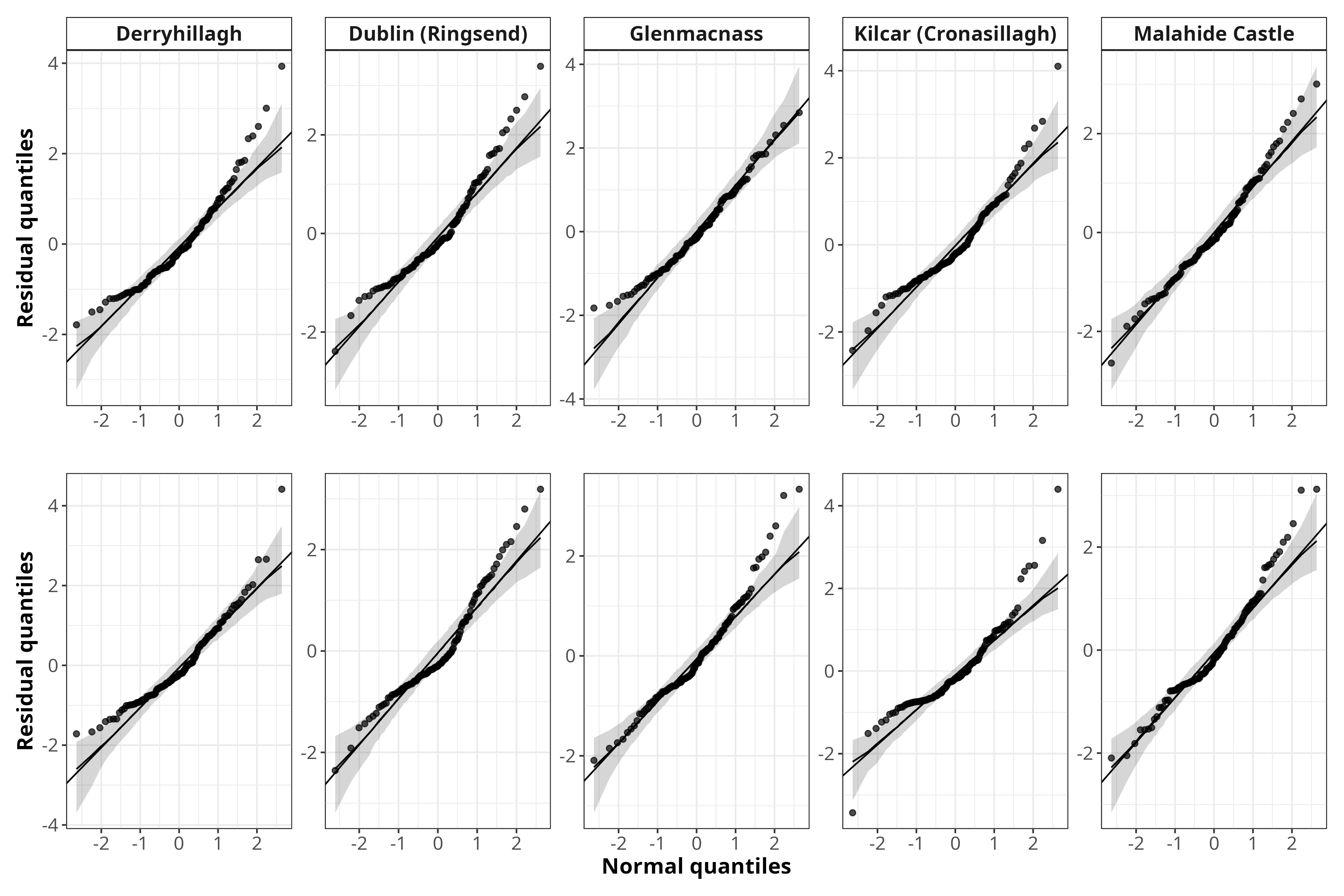}
    \caption{
        Normal quantile-quantile (QQ) plots of the estimated residuals for both conditional directions: precipitation conditional on wind speed (top) and wind speed conditional on precipitation (bottom).
        The solid black line denotes the quartile-based QQ reference line, while the shaded region represents pointwise 95\% simulation envelopes obtained from Gaussian samples.
        Results are shown for five sites: Malahide Castle, Dublin; Ringsend, Dublin; Glenmacnass, Wicklow; Kilcar, Donegal; and Derryhillagh, Mayo.
    }
    \label{fig:diag_app}
\end{figure}

To assess the adequacy of the Gaussian residual assumption underlying the proposed divergence measure, we examined normal QQ plots of the estimated margins of the residuals for several representative locations.
Figure~\ref{fig:diag_app} shows normal QQ plots of the standardised residuals for both conditioning directions at the five locations of interest identified in Figure~\ref{fig:chi} of the main manuscript.

Overall, the estimated residual distributions are broadly consistent with the working Gaussianity assumption, although some departures from normality are observed in the tails, as expected when using asymptotic models at finite thresholds \citep{Heffernan2004, Keef2013}.
No severe systematic deviations from Gaussianity are evident in Figure~\ref{fig:diag_app}.
Together with the strong empirical performance of our extremal clustering method observed in the simulation study, these diagnostics suggest that our proposed clustering methodology is reasonably robust to misspecification of the residual distribution.

\clearpage

\section{Elbow plots}

\begin{figure}[ht]
    \centering
    \includegraphics[width = 0.9\linewidth]{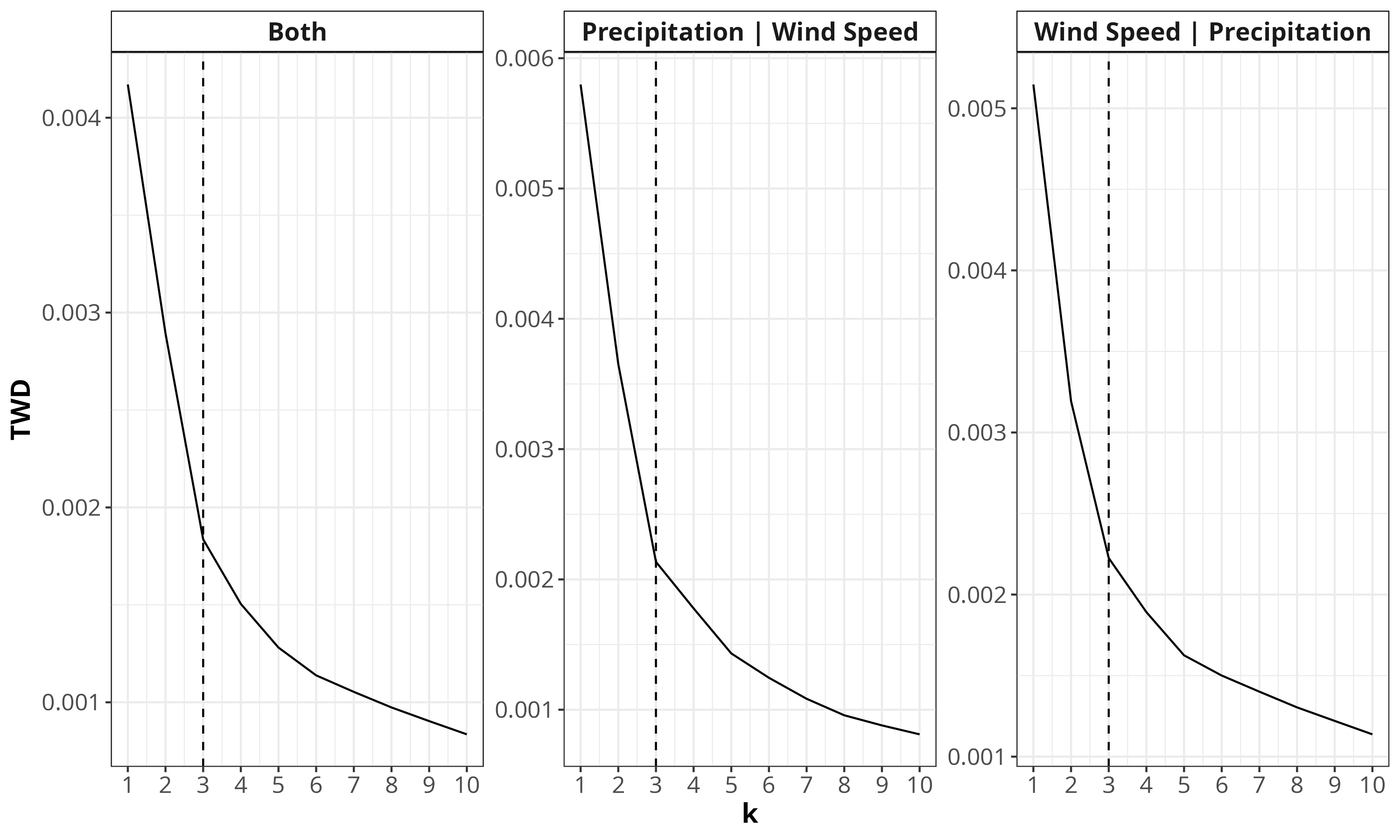}
    \caption{
      ``Elbow'' plots of the total within-group dissimilarity ($\mathrm{TWD}$) against the number of clusters $k$ for the estimated clusters obtained using the dissimilarity matrix for wind speed conditional on precipitation (centre), precipitation conditional on wind speed (right), and the combined dissimilarity matrix (left).
    }
    \label{fig:elbow_plots}
\end{figure}

Figure~\ref{fig:elbow_plots} gives the ``elbow'' plots of the total within-group dissimilarity ($\mathrm{TWD}$) against the number of clusters $k$, as described in Section~\ref{subsec:clustering} of the main text, for the motivating data application in Section~\ref{sec:application}.
These are shown for the estimated clusters obtained using the dissimilarity matrix for wind speed conditional on precipitation, precipitation conditional on wind speed, and the aggregated dissimilarity matrix $M$ as described in Equation~\eqref{eq:aggregate_matrix} of the main text.

\clearpage

\section{Sensitivity to overlapping ERA5 grid cells}
\label{sec:overlapping}

As noted in Section~\ref{subsec:app_data}, the ERA5 wind speed data are available on a relatively coarse $0.25^{\circ} \times 0.25^{\circ}$ grid, and so multiple Met Éireann gauges may fall within the same ERA5 grid cell and therefore share identical wind speed time series.
In our dataset, this occurs for 23 of the 59 locations, corresponding to 10 grid cells, including three cases where three locations share the same grid cell (but no more than three).

\begin{figure}[ht]
    \centering
    \includegraphics[width = 0.8\linewidth]{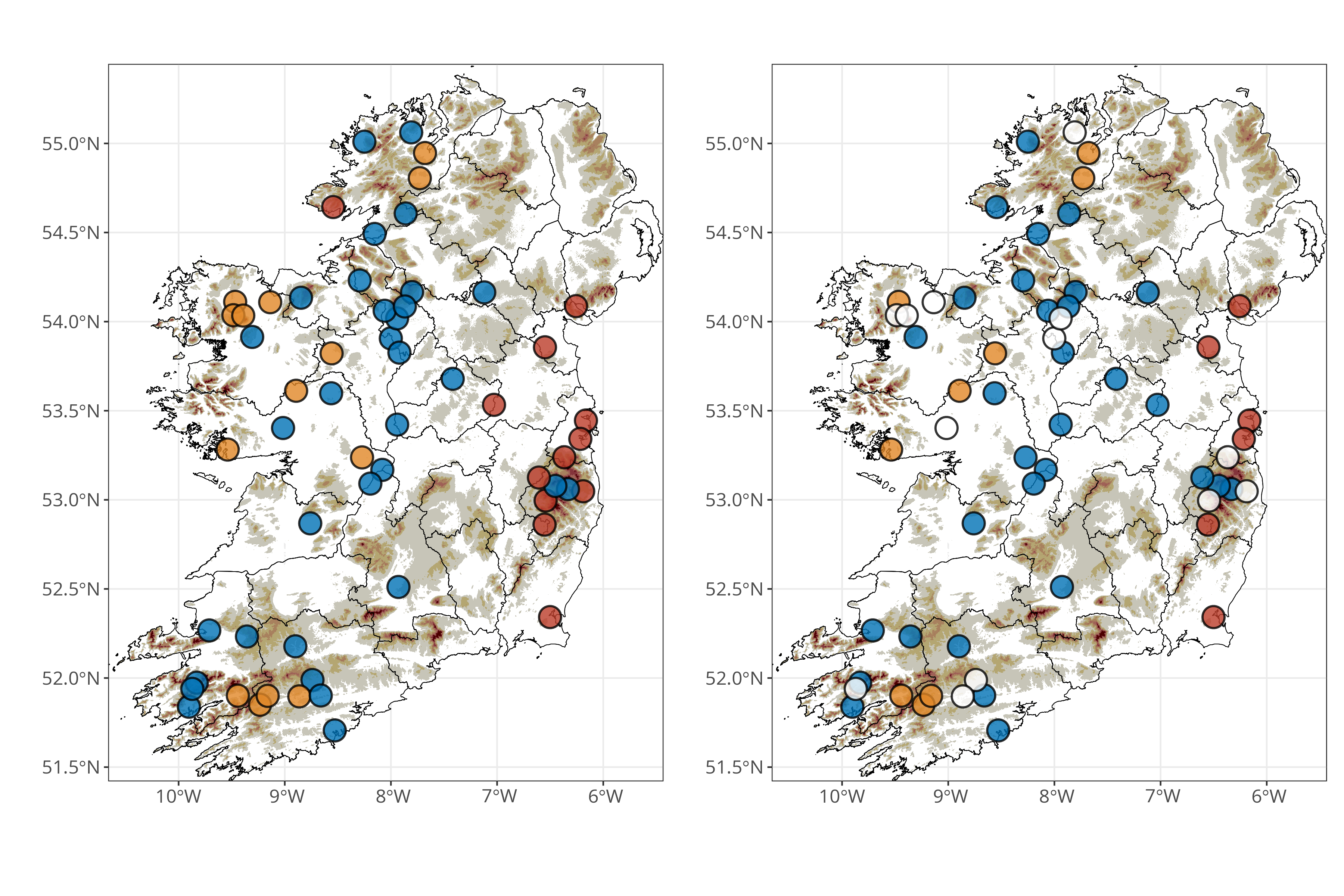}
    \caption{
    Estimated extremal clusters with the conditional extremes model parameters estimated at quantile level $q = 0.85$, using the aggregated dissimilarity matrix, for all 59 sites (left) and after retaining only one representative site from each overlapping ERA5 grid cell group, leaving 46 sites in total (right).
    Sites are coloured by cluster label, with removed sites in the right panel coloured in white. 
    }
    \label{fig:overlapping_sites}
\end{figure}

To assess the impact of this overlap on the estimated extremal clustering results, we performed a sensitivity analysis in which only one randomly selected representative site from each overlapping group was retained.
The resulting clustering solution, shown in Figure~\ref{fig:overlapping_sites}, exhibits very strong agreement with that obtained using all 59 sites, shown in Figure~\ref{fig:clust_sol} of the main text, with an ARI of 0.834 between the two solutions.
Moreover, the spatial clustering structures are visually very similar, with only 4 sites changing cluster assignment, all of which are located near cluster boundaries.
This suggests that the presence of overlapping ERA5 grid cells does not materially affect the substantive conclusions of the analysis in the main text. 

\clearpage

\section{Estimates for $k=2$ and $k=4$ clusters}

\begin{figure}[ht]
    \centering
    \includegraphics[width = 0.9\linewidth]{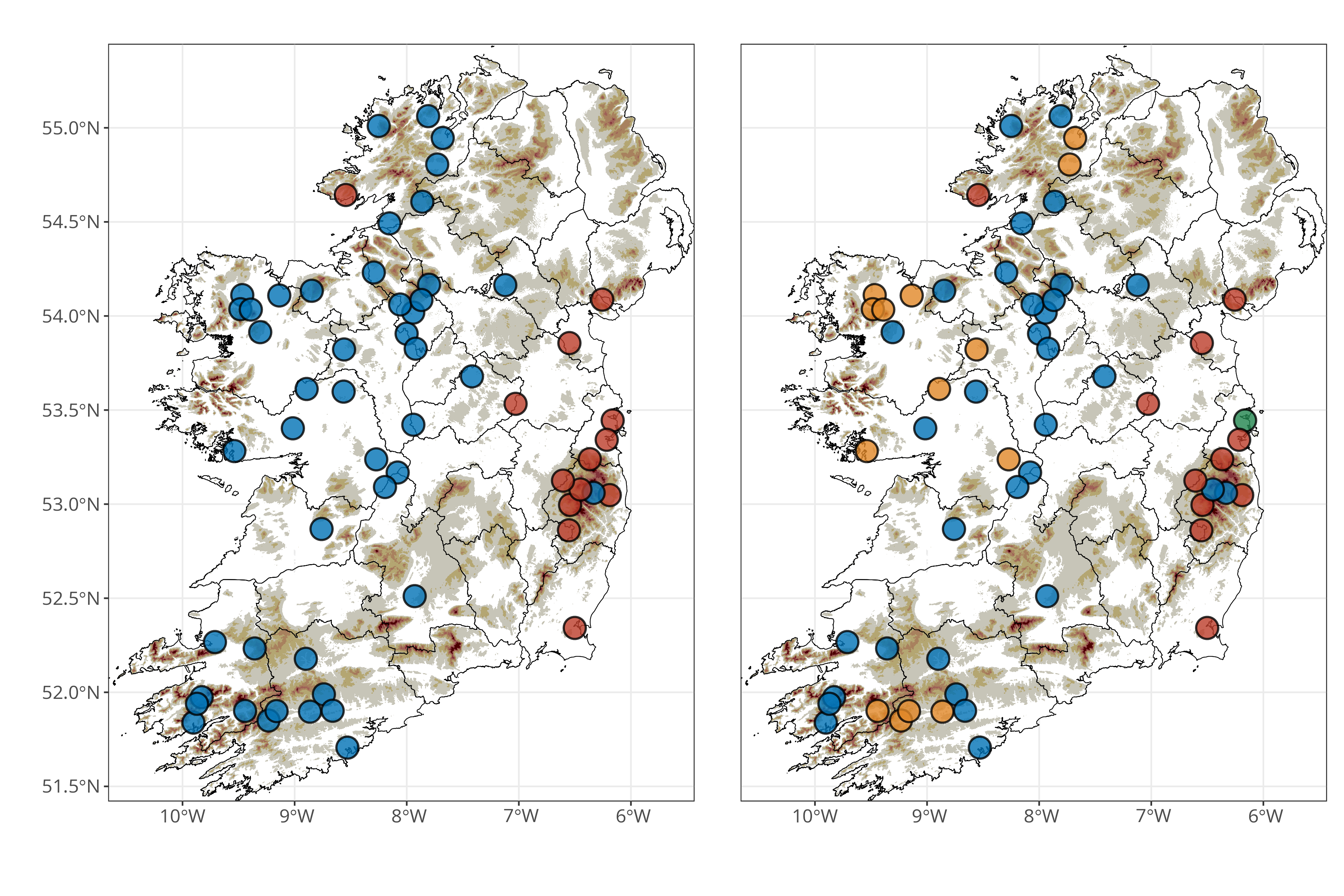}
    \caption{
    Extremal cluster estimates using the combined dissimilarity matrix for $k = 2$ (left) and $k=4$ (right) clusters. 
    Sites are coloured by cluster label.
    }
    \label{fig:cluster_dqu_k_2_4}
\end{figure}

Figure~\ref{fig:cluster_dqu_k_2_4} shows the estimates for $k=2$ and $k=4$ clusters, obtained using the aggregated dissimilarity matrix $M$, as mentioned in Section~\ref{subsec:app_clust} of the main text.

\clearpage

\section{Application comparison with \citet{Vignotto2021} method}
\label{sec:vignotto_compare}

\begin{figure}[ht]
    \centering
    \includegraphics[width = 0.9\linewidth]{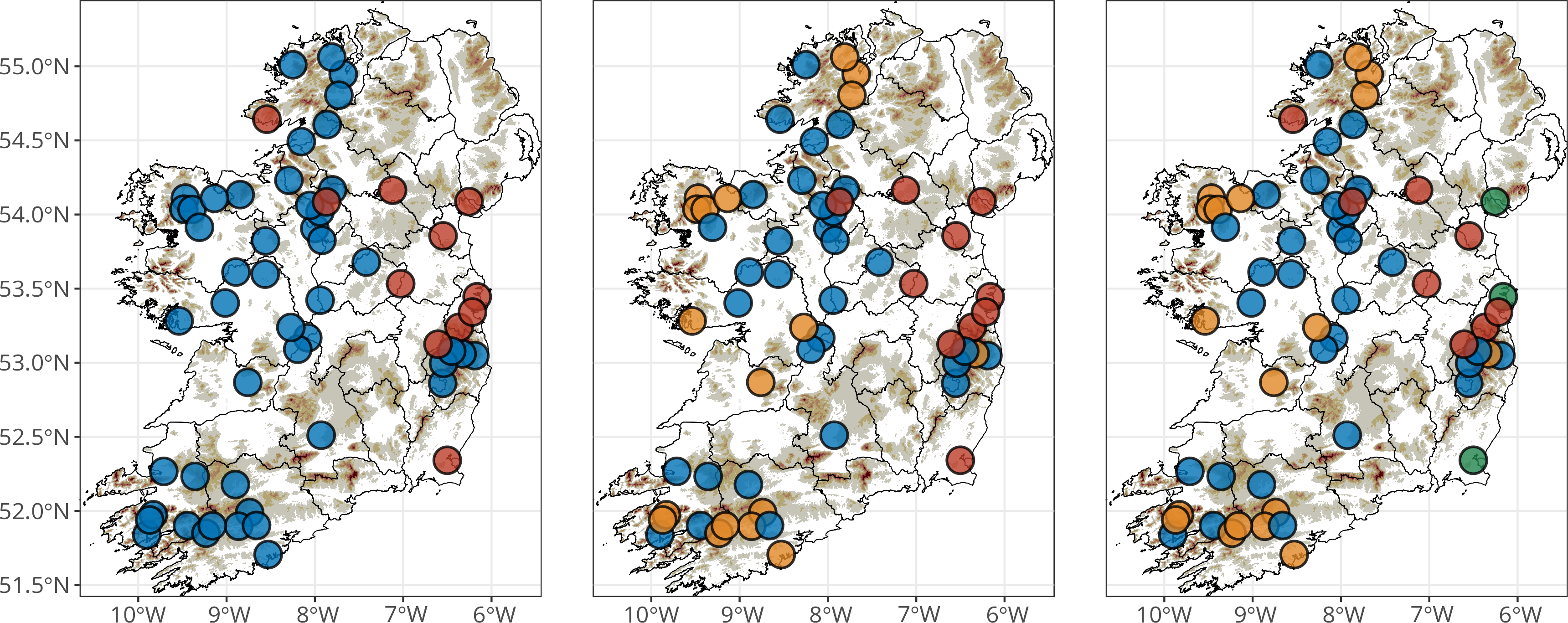}
    \caption{
      Extremal cluster estimates using the \citet{Vignotto2021} method for $k = 2$ (left) and $k=4$ (right) clusters. 
      Sites are coloured by cluster label.
    }
    \label{fig:vig_k_2_4}
\end{figure}

To compare the proposed clustering methodology with that of \citet{Vignotto2021}, we apply both approaches to the Irish meteorological dataset described in Section~\ref{sec:application}.
Figure~\ref{fig:vig_k_2_4} shows the clustering solutions obtained using the \citet{Vignotto2021} method for different numbers of clusters.

Overall, the clustering solutions obtained using the two extremal clustering methods have qualitatively similar large-scale spatial structure, particularly in the separation between eastern, central and western regions of Ireland.
The strongest agreement between the methods is observed for the two-cluster solution (ARI $\approx 0.58$), suggesting a relatively stable broad-scale partitioning of sites.
However, agreement weakens for the three- and four-cluster solutions (ARI $\approx 0.31$-$0.33$), indicating greater uncertainty in finer-scale cluster substructure and increased sensitivity to the choice of clustering methodology.
In particular, the four-cluster solutions include very small clusters, suggesting instability in the identification of minor subgroups.

\begin{figure}[ht]
    \centering
    \includegraphics[width = 0.9\linewidth]{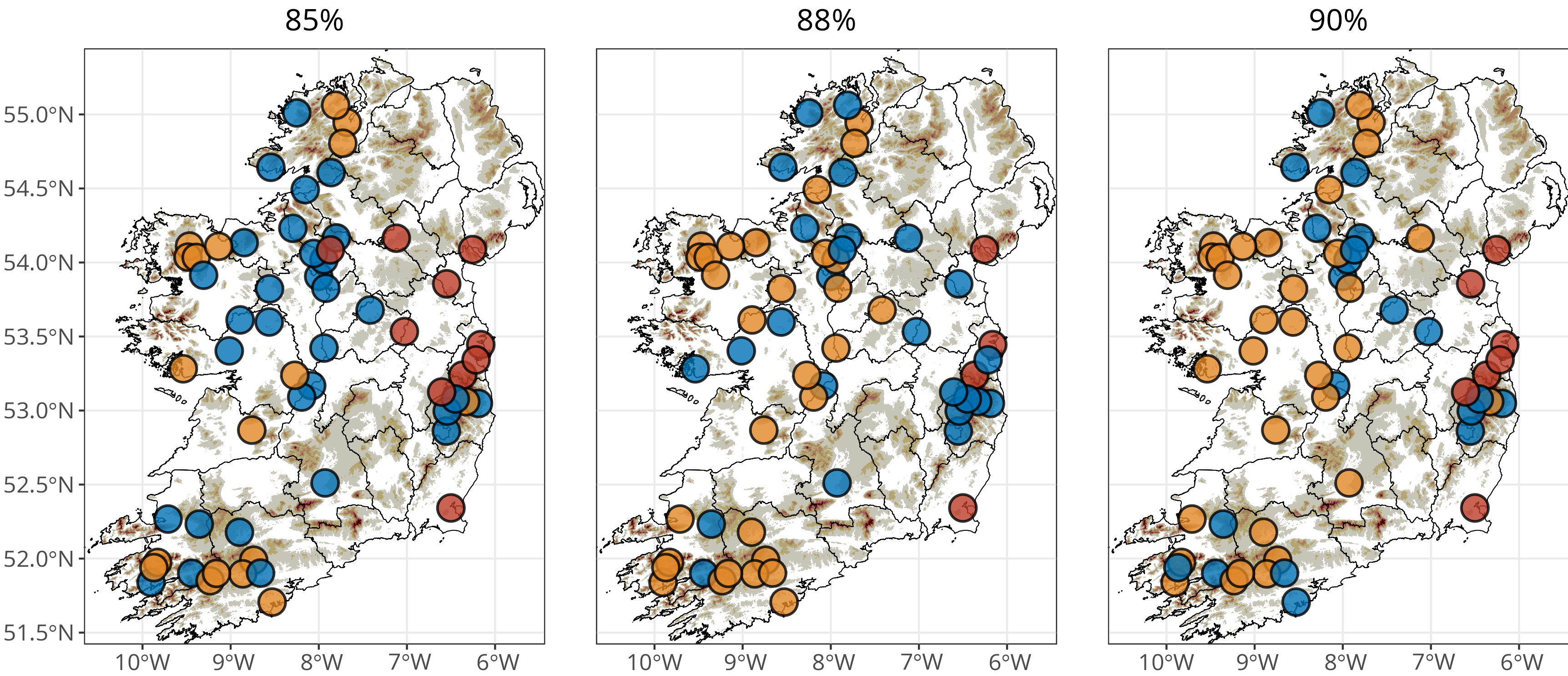}
    \caption{
      Estimated extremal clusters using the \citet{Vignotto2021} method estimated at $q = 0.85, q = 0.88$ and $q = 0.9$ quantiles. 
      Sites are coloured by cluster label. 
    }
    \label{fig:vig_thresh}
\end{figure}

We additionally assess sensitivity of the \citet{Vignotto2021} clustering approach to the choice of extremal threshold level.
Following \citet{Vignotto2021}, we vary the quantile level $q$ used to define joint extreme events in the transformed Pareto scale representation (see Section~2.3 of \citet{Vignotto2021}).
The resulting clustering estimates for the three-cluster solution are shown in Figure~\ref{fig:vig_thresh}, for $q \in \{0.85, 0.88, 0.90\}$.

Compared with the proposed CE-based clustering methodology, the \citet{Vignotto2021} approach exhibits greater variability across threshold choices.
Pairwise agreement between clustering solutions across thresholds is substantially lower for the \citet{Vignotto2021} method (ARI $= 0.16$--$0.33$) than for our proposed CE-based approach (ARI $= 0.51$--$0.59$), suggesting increased sensitivity to threshold specification in this application.

\clearpage

\clearpage
